\def\FINALSUBMISSION{}%
\documentclass{article}
\pdfinfoomitdate=1 \pdftrailerid{} \pdfsuppressptexinfo=-1
\usepackage{microtype}
\usepackage{graphicx}
\usepackage{booktabs}
\usepackage{longtable}
\usepackage{multirow}
\usepackage{array}
\usepackage{amsmath,amssymb}
\usepackage{xcolor}
\usepackage{tikz}
\usetikzlibrary{positioning,arrows.meta}
\usepackage{xurl}
\usepackage{needspace}
\usepackage[hypertexnames=false]{hyperref}
\newcommand{\papertitle}{Sandman: A Sandbox Service for Production Agentic Reinforcement Learning}

\usepackage[accepted]{mlsys2025}
\usepackage{float}
\usepackage{placeins}
\makeatletter
\renewcommand{\Notice@String}{Preprint.}%
\makeatother
\usepackage{xparse}
\newif\iffinalsubmission
\ifdefined\FINALSUBMISSION\finalsubmissiontrue\fi
\NewDocumentCommand{\TODO}{m}{\iffinalsubmission\PackageError{aar}{TODO left: \detokenize{#1}}{}\fi
  \typeout{TODO: \detokenize{#1}}\textcolor{red}{\textbf{[TODO: #1]}}}
\def\coderepo{}
\newcommand{\codestatus}{\ifx\coderepo\empty, to be published\else, available at \url{\coderepo}\fi}
\newcommand{\sys}{Sandman}
\newcommand{\execd}{\texttt{execd}}

\newcommand{\appref}[1]{\ref*{#1}}

\mlsystitlerunning{Sandman: A Sandbox Service for Production Agentic RL}
\hypersetup{pdfsubject={}, pdftitle={\papertitle}, pdfauthor={Yiming Yan, Hongyang Wang, Wenbo Chen, Xuan Wang, Yichuan Li, Jingguo Liu, Xian Li, Leo Fan, Chen Qi, Siwei Li, Qianli Wu, Ruochen Wen, Qian Qu, Zheshen Wang, Benoit Dumoulin},
  pdfkeywords={Agentic reinforcement learning, sandboxes, Kubernetes, ML infrastructure, operations}}
\begin{document}
\twocolumn[
\mlsystitle{Sandman: A Sandbox Service for Production\texorpdfstring{\\}{ }Agentic Reinforcement Learning}
\mlsyssetsymbol{equal}{*}
\begin{mlsysauthorlist}
\mlsysauthor{Yiming Yan}{equal,amzn}
\mlsysauthor{Hongyang Wang}{equal,amzn}
\mlsysauthor{Wenbo Chen}{amzn}
\mlsysauthor{Xuan Wang}{amzn}
\mlsysauthor{Yichuan Li}{amzn}
\mlsysauthor{Jingguo Liu}{amzn}
\mlsysauthor{Xian Li}{amzn}
\mlsysauthor{Leo Fan}{amzn}
\mlsysauthor{Chen Qi}{amzn}
\mlsysauthor{Siwei Li}{amzn}
\mlsysauthor{Qianli Wu}{amzn}
\mlsysauthor{Ruochen Wen}{amzn}
\mlsysauthor{Qian Qu}{amzn}
\mlsysauthor{Zheshen Wang}{amzn}
\mlsysauthor{Benoit Dumoulin}{amzn}
\end{mlsysauthorlist}
\mlsysaffiliation{amzn}{Amazon}
\mlsyscorrespondingauthor{Yiming Yan}{\mbox{yyming009@gmail.com}}
\mlsyskeywords{Agentic reinforcement learning, sandboxes, Kubernetes, ML infrastructure, operations}
\vskip 0.3in
\begin{abstract}
Agentic reinforcement learning (RL) needs stateful environments for tool execution and evaluation.
Training jobs request thousands of environments in bursts that can overwhelm shared cluster services, while
cancelled or timed-out requests can leave resources running.
We present Sandman, a sandbox service built and operated at Amazon for large-scale agentic RL workloads in
production on shared Kubernetes.
A small daemon mounted into unmodified task images preserves files and processes across calls.
Sandman paces creation requests, checks sandbox readiness, and retains cleanup responsibility after callers time
out or cancel.
Production deployments created 3.1 million sandboxes for 122 coding-agent training runs in two weeks.
Peak production concurrency was 11,858 sandboxes.
In one 10,000-request multi-client burst, Sandman made half the sandboxes usable in 116 seconds and all in 308 seconds.
Full delivery took 44\% less time than a tuned configuration of agent-sandbox, an open-source sandbox manager for
Kubernetes. Sandman left no sandboxes at client exit.
On software-engineering task images, median no-op command latency was 2.0 ms, compared with 14.3 ms for
agent-sandbox and 203.5 ms for OpenSandbox.
We also measure provisioning costs and sandbox delays within training runs, and report design trade-offs and
operational lessons from production use and comparisons with three open-source systems.
We will open-source Sandman's daemon, provider and pool.
\end{abstract}

]
\printAffiliationsAndNotice{\mlsysEqualContribution}
\hypersetup{pdftitle={\papertitle}, pdfauthor={Yiming Yan, Hongyang Wang, Wenbo Chen, Xuan Wang, Yichuan Li, Jingguo Liu, Xian Li, Leo Fan, Chen Qi, Siwei Li, Qianli Wu, Ruochen Wen, Qian Qu, Zheshen Wang, Benoit Dumoulin}}
\section{Introduction}
\label{sec:intro}

\begin{figure*}[t]
\centering
\includegraphics[width=\textwidth]{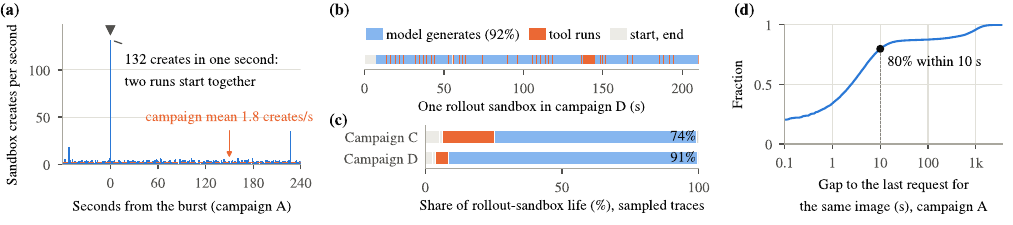}
\caption{Production demand arrives in bursts and repeatedly uses the same task images. (a)~Creates per second around
campaign~A's largest burst, when two runs started together. (b)~A campaign-D rollout with median sandbox lifetime and
tool-call count. (c)~Time spent in each phase of a rollout sandbox's life, from sampled traces of campaigns C and D.
(d)~Cumulative distribution of intervals between requests for the same image within a campaign-A run.}
\label{fig:motiv}
\end{figure*}

In agentic reinforcement learning (RL), parallel rollouts alternate model inference with commands and file edits.
Their tools need task-specific software and environments that preserve files and processes between calls.
Training jobs repeatedly create and release these environments for tool execution and evaluation.

Concurrent requests can overload shared admission and network setup even when compute capacity is available.
Prewarming shortens the wait for an environment but reserves resources before requests arrive. An accepted request
can still start an environment after its caller cancels or times out.

We present Sandman, a sandbox service built and operated at Amazon for multiple production agentic RL workloads
on a shared Amazon Elastic Kubernetes Service (EKS) cluster. Frameworks supply task images and resource
requirements; Sandman manages creation, readiness and cleanup (Figure~\ref{fig:arch}).

A small daemon mounted into each unmodified task image provides command and file operations. The provider paces
creates, separates Kubernetes API calls from daemon checks, and retains cleanup responsibility after cancellation.
Creation is on demand by default, with optional prewarming and checked reuse for grading.

In two weeks, production deployments created 3.1 million sandboxes across 122 coding-agent training runs, with up
to 432 B200 GPUs active at once (\S\ref{sec:evalprod}). We use production traces to characterize demand
(Figure~\ref{fig:motiv}; \S\ref{sec:motivation}) and evaluate Sandman primarily on software-engineering (SWE) workloads.

This paper makes three contributions:
\begin{enumerate}
\item We describe Sandman's design and implementation for production tool execution and evaluation, and will
open-source its daemon, provider and pool (\S\S\ref{sec:system}--\ref{sec:design}).
\item We compare Sandman with three open-source systems across burst delivery, tool-call latency and cleanup.
In a 10,000-request burst with 16 client processes per system, Sandman completed delivery in 308 seconds, compared
with 547 seconds for a tuned agent-sandbox configuration. On SWE task images, median no-op command latency was
2.0 ms, versus 14.3 ms for agent-sandbox and 203.5 ms for OpenSandbox. Sandman also left no sandboxes at client
exit in a 300-create cancellation test (\S\ref{sec:evalcmp}).
\item Production traces, paired training experiments and failure investigations document provisioning costs,
sandbox waits in training, and the operational lessons that shaped the service
(\S\S\ref{sec:evalprep}--\ref{sec:lessons}).
\end{enumerate}

\section{Background and Motivation}
\label{sec:motivation}

\subsection{Rollouts and sandboxes in agentic RL}
\label{sec:workload}

Each SWE task provides an issue, a repository at a specified base commit, a task image and tests. The agent runs
commands and edits files over multiple turns in a sandbox created from that image.

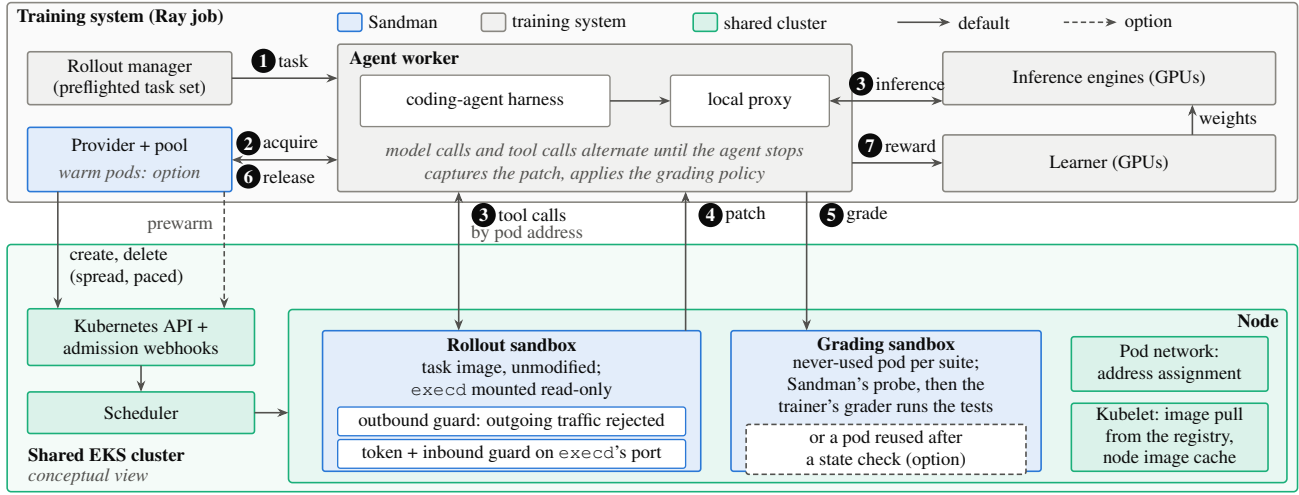
\begin{figure*}[t]
\centering
\providecommand{\execd}{\texttt{execd}}%
\definecolor{aarBlue}{HTML}{2A78D6}\definecolor{aarAqua}{HTML}{1BAF7A}%
\definecolor{aarInk}{HTML}{0B0B0B}\definecolor{aarInkTwo}{HTML}{52514E}\definecolor{aarGray}{HTML}{898781}%
\begin{tikzpicture}[x=1mm, y=1mm, font=\fontsize{7}{8.2}\selectfont, line join=round,
  aarours/.style={draw=aarBlue, fill=aarBlue!13, line width=0.6pt, rounded corners=1pt},
  aarshared/.style={draw=aarAqua, fill=aarAqua!16, line width=0.6pt, rounded corners=1pt},
  aaroursregion/.style={draw=aarBlue, fill=aarBlue!4, line width=0.6pt, rounded corners=1.5pt},
  aartrain/.style={draw=aarGray, fill=aarGray!12, line width=0.6pt, rounded corners=1pt},
  aartrainregion/.style={draw=aarGray, fill=aarGray!4, line width=0.6pt, rounded corners=1.5pt},
  aartraininner/.style={draw=aarGray, fill=white, line width=0.5pt, rounded corners=1pt},
  aarsharedregion/.style={draw=aarAqua, fill=aarAqua!5, line width=0.6pt, rounded corners=1.5pt},
  aarnoderegion/.style={draw=aarAqua, fill=aarAqua!10, line width=0.6pt, rounded corners=1.5pt},
  aaropt/.style={draw=aarInkTwo, line width=0.6pt, dash pattern=on 2pt off 1.2pt, rounded corners=1pt},
  aarinner/.style={draw=aarBlue, fill=white, line width=0.5pt, rounded corners=1pt},
  aartxt/.style={align=center, inner sep=0pt, text=aarInk},
  aartitle/.style={inner sep=0pt, text=aarInk, font=\fontsize{7}{8.2}\selectfont\bfseries},
  aarflow/.style={-{Stealth[length=1.5mm, width=1.15mm]}, line width=0.6pt, draw=aarInkTwo},
  aarboth/.style={{Stealth[length=1.5mm, width=1.15mm]}-{Stealth[length=1.5mm, width=1.15mm]},
    line width=0.6pt, draw=aarInkTwo},
  aarcur/.style={{Stealth[length=1.5mm, width=1.15mm]}-{Stealth[length=1.5mm, width=1.15mm]},
    line width=0.6pt, draw=aarInkTwo, dash pattern=on 3pt off 1pt on 0.6pt off 1pt},
  aaroptflow/.style={-{Stealth[length=1.5mm, width=1.15mm]}, line width=0.6pt, draw=aarInkTwo,
    dash pattern=on 2pt off 1.2pt},
  aarnum/.style={circle, fill=aarInk, text=white, inner sep=0pt, minimum size=3.2mm,
    font=\fontsize{7}{7}\selectfont\bfseries}]
\draw[aartrainregion] (0.3,38.6) rectangle (170.9,65.0);
\node[aartitle, anchor=north west] at (1.5,64.2) {Training system (Ray job)};
\draw[aarours] (44,61.2) rectangle (47.2,63.6);   \node[aartxt, anchor=west] at (48.0,62.4) {Sandman};
\draw[aartrain] (63,61.2) rectangle (66.2,63.6);  \node[aartxt, anchor=west] at (67.0,62.4) {training system};
\draw[aarshared] (91,61.2) rectangle (94.2,63.6); \node[aartxt, anchor=west] at (95.0,62.4) {shared cluster};
\draw[aarflow] (118,62.4) -- (125,62.4);           \node[aartxt, anchor=west] at (126,62.4) {default};
\draw[aaroptflow] (140,62.4) -- (147,62.4);        \node[aartxt, anchor=west] at (148,62.4) {option};
\draw[aartrain] (3,51.5) rectangle (30,58.5);   \node[aartxt] at (16.5,55.0) {Rollout manager\\(preflighted task set)};
\draw[aarours] (3,40.0) rectangle (30,48.5);
\node[aartxt] at (16.5,45.9) {Provider + pool};
\node[aartxt, text=aarInkTwo] at (16.5,42.4) {\itshape warm pods: option};
\draw[aartrain] (44,40.0) rectangle (112,59.5);
\node[aartitle, anchor=north west] at (45.5,58.7) {Agent worker};
\draw[aartraininner] (47,48.6) rectangle (80,55.4);  \node[aartxt] at (63.5,52.0) {coding-agent harness};
\draw[aartraininner] (88,48.6) rectangle (109,55.4); \node[aartxt] at (98.5,52.0) {local proxy};
\draw[aarflow] (80,52.0) -- (88,52.0);
\node[aartxt, text=aarInkTwo] at (78,45.3) {\itshape model calls and tool calls alternate until the agent stops};
\node[aartxt, text=aarInkTwo] at (78,42.3) {\itshape captures the patch, applies the grading policy};
\draw[aartrain] (124,51.5) rectangle (168,58.5); \node[aartxt] at (146,55.0) {Inference engines (GPUs)};
\draw[aartrain] (124,40.0) rectangle (168,47.5); \node[aartxt] at (146,43.75) {Learner (GPUs)};
\draw[aarflow] (30,55.0) -- (44,55.0);
\node[aarnum] at (34.2,57.3) {1}; \node[aartxt, anchor=west] at (36.1,57.3) {task};
\draw[aarboth] (30,44.2) -- (44,44.2);
\node[aarnum] at (32.2,46.4) {2}; \node[aartxt, anchor=west] at (34.1,46.4) {acquire};
\node[aarnum] at (32.2,42.0) {6}; \node[aartxt, anchor=west] at (34.1,42.0) {release};
\draw[aarboth] (109,52.0) -- (124,52.0);
\node[aarnum] at (113.2,54.3) {3}; \node[aartxt, anchor=west] at (115.1,54.3) {inference};
\draw[aarflow] (112,43.8) -- (124,43.8);
\node[aarnum] at (114.4,46.1) {7}; \node[aartxt, anchor=west] at (116.3,46.1) {reward};
\draw[aarflow] (157,47.5) -- (157,51.5) node[midway, right=0.8mm, aartxt] {weights};
\draw[aarsharedregion] (0.3,0.3) rectangle (170.9,33.0);
\node[aartitle, anchor=south west] at (3,4.3) {Shared EKS cluster};
\node[aartxt, anchor=south west, text=aarInkTwo] at (3,1.2) {\itshape conceptual view};
\draw[aarshared] (3,17.0) rectangle (33,24.5); \node[aartxt] at (18,20.75) {Kubernetes API +\\admission webhooks};
\draw[aarshared] (3,8.0) rectangle (33,13.5);  \node[aartxt] at (18,10.75) {Scheduler};
\draw[aarflow] (18,17.0) -- (18,13.5);
\draw[aarflow] (7,40.0) -- (7,24.5);
\draw[aaroptflow] (29,40.0) -- (29,24.5);
\node[aartxt, anchor=west] at (8.5,31.6) {create, delete};
\node[aartxt, anchor=west] at (8.5,28.6) {(spread, paced)};
\node[aartxt, anchor=east, text=aarInkTwo] at (27.6,35.4) {prewarm};
\draw[aarnoderegion] (37.5,1.5) rectangle (169.5,24.4);
\node[aartitle, anchor=north east] at (168.5,23.8) {Node};
\draw[aarflow] (33,10.75) -- (37.5,10.75);
\draw[aarours] (42,3.0) rectangle (92,21.6);
\node[aartitle] at (67,19.7) {Rollout sandbox};
\node[aartxt] at (67,15.1) {task image, unmodified;\\\execd{} mounted read-only};
\node[aartxt, aarinner, minimum width=46mm, minimum height=3.8mm] at (67,9.6)
  {outbound guard: outgoing traffic rejected};
\node[aartxt, aarinner, minimum width=46mm, minimum height=3.8mm] at (67,5.3)
  {token + inbound guard on \execd{}'s port};
\draw[aarours] (96,3.0) rectangle (137,21.6);
\node[aartitle] at (116.5,19.7) {Grading sandbox};
\node[aartxt] at (116.5,14.3) {never-used pod per suite;\\Sandman's probe, then the\\trainer's grader runs the tests};
\node[aartxt, aaropt, fill=white, minimum width=37mm, minimum height=6.6mm] at (116.5,5.9)
  {or a pod reused after\\a state check (option)};
\draw[aarshared] (141,13.6) rectangle (167,20.8); \node[aartxt] at (154,17.2) {Pod network:\\address assignment};
\draw[aarshared] (141,3.0) rectangle (167,12.0);
\node[aartxt] at (154,7.5) {Kubelet:\ image pull\\from the registry,\\node image cache};
\draw[aarboth] (60,40.0) -- (60,21.6);
\node[aarnum] at (63.3,36.9) {3}; \node[aartxt, anchor=west] at (65.2,36.9) {tool calls};
\node[aartxt, anchor=west, text=aarInkTwo] at (61.6,34.5) {by pod address};
\draw[aarflow] (90,21.6) -- (90,40.0);
\node[aarnum] at (93.3,36.9) {4}; \node[aartxt, anchor=west] at (95.2,36.9) {patch};
\draw[aarflow] (106,40.0) -- (106,21.6);
\node[aarnum] at (109.3,36.9) {5}; \node[aartxt, anchor=west] at (111.2,36.9) {grade};
\end{tikzpicture}%
\caption{Sandman's components and a conceptual SWE rollout, colored by code ownership. The provider and pool run
inside the training job, which captures patches, runs graders and applies the grading policy. Our SWE defaults apply
both network guards to every sandbox. An agent worker receives a task (1), acquires a sandbox (2), alternates model
inference and tool calls (3), extracts a patch (4), grades it in a separate pod (5), releases both sandboxes (6) and
sends the reward to the learner (7).}
\label{fig:arch}
\end{figure*}

When the agent stops, the trainer extracts a patch against the task's base commit and applies it in a separate
grading sandbox from the same image. Tests produce a reward for group relative policy optimization
(GRPO)~\cite{deepseekmath}. The trainer then releases the environments (Figure~\ref{fig:arch}).

Training is asynchronous: the learner updates the model from completed, graded groups while other rollouts
continue. This overlap can hide sandbox delays, but a waiting rollout still occupies an in-flight slot.
Slow acquisition can therefore reduce the rate of completed rollouts.

Our workload measurements come from 122 production training runs on SWE-bench Verified~\cite{swebench,swebenchverified},
grouped into four campaigns, A--D (Table~\appref{tab:a-campaigns}). Some controlled experiments use
SWE-Gym~\cite{swegym}. Sandman also serves other production agentic RL workloads.

We call a rollout record a \emph{trajectory} and an agent-session start an \emph{attempt}. An \emph{acquisition}
spans a trainer request through delivery of a usable sandbox. It may include a \emph{create}, which ends when the daemon answers.

\subsection{Motivation}
\label{sec:scenarios}

Our production runs averaged about 26,000 sandboxes each, but demand was uneven. One campaign-A run started
64 trajectories within a second. When two runs started together, creation requests reached 132 in one second,
compared with a campaign mean of 1.8 per second (Figure~\ref{fig:motiv}a). Each create passes through shared admission,
scheduling and network setup, whose capacity can limit creation even when compute is available.

A rollout sandbox remains allocated while the model generates its next action. In sampled traces from campaigns
C and D, this waiting occupied 74--91\% of sandbox lifetime (Figure~\ref{fig:motiv}c). Preserving files and process
state keeps sandbox resources allocated between tool calls.

A run used 24--88 task images, and about 80\% of requests for the same image within a campaign-A run arrived less
than ten seconds apart (Figure~\ref{fig:motiv}d). These repeated requests offer an opportunity to prewarm
environments. A warm pod shortens acquisition but reserves CPU and memory, and cannot serve a different image.
Its value depends on demand and on how acquisition delays affect training progress.

Among 500 SWE-bench Verified reference patches, 15 received different verdicts across execution conditions, including
11 when outbound traffic was denied. The same code can therefore receive different rewards. Grading requires
explicit execution conditions and checks that tests behave as intended under that configuration
(\S\ref{sec:grading}).

\section{Sandman Overview}
\label{sec:system}

\subsection{Components and the path of a rollout}
\label{sec:components}

Sandman comprises a pod provider, tool daemon and optional pool (Figure~\ref{fig:arch}; Table~\appref{tab:api}).

\execd{} is an HTTP daemon mounted read-only into the unmodified task image. It needs only bash and Python~3.
A persistent bash process preserves the working directory, environment and background processes across calls.

The provider creates one pod from an image, CPU, memory and workspace specification. CPU and memory requests equal
their limits. Pods carry no service-account token. Before handoff, the readiness probe and a provider \execd{} health
check must pass within the job's time budget, including image pulling.

The pool is a Ray~\cite{ray} actor shared by the trainer's workers. It can keep unused pods for each (image,
base commit), cap live pods and concurrent creates, and pace its creates under one budget.

Agent workers run outside the sandboxes. Each agent slot has a Ray worker whose tools call \execd{} at the pod's
address; model calls go through a local proxy.

\subsection{Responsibilities and the trust boundary}
\label{sec:ownership}

Sandman uses the cluster's API server, admission webhooks, DNS, node groups and pod network, provided by the Virtual
Private Cloud Container Network Interface plugin (VPC CNI). Our team operates these services for other workloads too.
Sandman's service account can create and delete pods in its namespace, but cannot read webhook configurations or
create network policies. A pool or worker owns each pod during use, and the provider deletes it.
Owner references let Kubernetes remove remaining sandboxes when the trainer's pod is deleted.

The trusted components are the trainer, graders and cluster control plane. Code generated by our models and pods
belonging to other workloads are outside that boundary. Model-generated code runs as root in runc containers that
share the node's kernel with other pods (\S\ref{sec:boundaries}).

\section{Design Considerations}
\label{sec:design}

\subsection{Where sandboxes run: managed service or self-managed}
\label{sec:where}

We first used Amazon Bedrock AgentCore Code Interpreter~\cite{agentcoreci}. It then offered one sandbox size
(2\,vCPUs, 8\,GB memory, 10\,GB disk) and no custom task images. Default per-account limits were
30 calls/s on session starts, stops and invokes~\cite{agentcorequotas}, and 1,000 concurrent sessions.
Campaign~D later needed up to 1,728 trajectories in flight across runs. We obtained higher start and invoke limits,
but when Python and math rollouts recycled sessions together, stop calls reached about 117/s including retries,
and trajectories timed out. We did not benchmark other hosted runtimes (\S\ref{sec:related}).

Sandman instead creates pods on the training cluster's CPU nodes. In mid-October, the cluster had 378 such nodes,
each with 96 vCPUs and local NVMe SSDs, across four families: c5d, m5d, m6id and m8id. The container image store uses NVMe for large task-image caches without filling the root disk. Jobs choose their
images, resources, readiness budget and network rules. Workers connect over the cluster network, without a privileged
Sandman daemon on each node. Creates pass through the shared API server, webhooks and pod network.

\subsection{Sandbox creation and placement}
\label{sec:traffic}

At 2\,vCPUs and 8\,GiB per sandbox, a c5d node fits 21 sandboxes by memory but 46 by CPU. Creates also depend on
the API server, admission webhooks, scheduler, address assignment and kubelet. The network plugin~\cite{vpccni}
preallocates IP addresses per node; its free-IP target is not an exact spare-address count, especially with prefix
delegation. Bursts concentrated on a few nodes stalled address assignment despite available cluster capacity
(Table~\appref{tab:c-limits}). The provider requests soft topology spreading with maximum skew 2, which the
scheduler may override.

Each provider uses a token bucket to limit the create rate and allowed burst. The pool shares one provider budget
across on-demand creates, warm builds and replacements. Without the pool, each worker paces independently, and no
budget spans jobs. Workers fall back to a direct create when the pool raises an error other than a timeout, even
when the pool has reached its pod cap. This fallback bypasses the pool's budget and cap and served 4.3\% of the pooled arm's gradings in the
pre-registered comparison.

The provider retries creates refused with HTTP 429 or a webhook timeout up to 30 times through the pacer, using
backoff and jitter~\cite{k8sapf}. Other errors follow the cleanup path in \S\ref{sec:cancel}. One watch monitors readiness. Kubernetes API calls use
12 threads; daemon calls have a separate pool and do not queue behind creates. Achieved rates depend on response
times and concurrent demand (settings: Table~\appref{tab:configuration}).

The trainer sends each pod address to the harness and waits for acknowledgement. Agents then connect directly.
Our earlier Service-per-slot design added an object per create and delayed connectivity while routes were programmed.
On warm pods, the median time from the acquisition request to the first successful command was 2.3\,s with
Services and 0.05\,s with direct pod addresses (Figure~\ref{fig:scale}c).
We disable Service links, which otherwise inject environment variables for eligible ClusterIP Services into new
containers~\cite{k8sserviceenv}.

Pacing adds client waiting, and spreading may require an image pull. Concentrating requests on cached-image nodes
instead accumulated address failures.
The scheduler's image-locality score also sees only the reported image list, which defaults to at most 50 images
per node~\cite{kubeletconfig}. Raising warm-address targets reserves more subnet capacity for bursts; larger bursts
may still require more nodes or slower creation.

\subsection{Provisioning: on demand, prewarmed or reused}
\label{sec:sources}

Image pulls account for much of the delay in on-demand creation (\S\ref{sec:evalprep}). Nodes cache task images on
local disks until disk pressure causes eviction. An admission rule requires a registry check for images specified
by tag, so updates such as security fixes can take effect. Digest references are pulled only when absent from the node. Both paths reuse
cached layers.

Sandman creates on demand by default. Warm pods are started but unused, reserving full resources while waiting. Handoff requires a health check and a
match on CPU, memory and workspace, then triggers a replacement create. The pool removes warm pods when their
image has not been requested within the idle timeout.

Grading pods can also be reused. The commands \texttt{git reset} and \texttt{git clean} do not restore the whole
environment. At creation, \execd{} records a baseline for state that graders were observed to change: the
repository, installed packages, system and home directories, environment, time zone and processes. After grading,
Sandman restores this tracked state or deletes the pod. Only the recorded baseline is checked (Appendix~\appref{app:i}). A pod is reused at most 8 times and retired if its verdict
contradicts an earlier grading.

Warm pods are keyed by the requested image reference. All campaigns and experiments used digests to keep rollout,
warm and grading images identical; mutable tags would not ensure this.

Each warm pod serves only its own image, so pool sizing depends on each image's request rate, burst size and
create time. We did not adopt lazy image loading because it requires another snapshotter on every node.
CPU and memory requests equal their limits throughout. We do not overcommit resources during idle periods.

\subsection{Cleanup when the caller gives up}
\label{sec:cancel}

If caller cancellation stops cleanup after the API server accepts a create, a pod can become ready without a
waiting caller while a retry creates another. In a stress test, 5,904 pods were removed only
by garbage collection at teardown. At the peak, there were at least 4,181 more live pods than agent slots. Table~\ref{tab:cleanup} lists this
and two other ways pods can outlive their callers.

The provider finishes creates after caller cancellation and runs cleanup on every exit, including a second cancellation. The pool retains the pending create when an acquisition
times out: a late pod without an agent label enters the warm pool; other late pods are deleted. Each create carries a
request identifier (nonce). After a transport or server error, the provider looks up the pod by name. It deletes
the pod only if the identifier matches. Unconfirmed deletes are retried until the job drains. Pending deletes are
lost if the provider's event loop closes. Kubernetes can remove these pods through their owner references when the
trainer pod is deleted.

\begin{table}[t]
\caption{Three ways a pod can outlive its caller and how Sandman handles them. Fault-injection results:
Table~\appref{tab:f-windows}; all 25 cluster trials left no pod (Appendix~\appref{app:f}).}
\label{tab:cleanup}
\centering\small
\begin{tabular}{@{}>{\raggedright\arraybackslash}p{2.1cm}>{\raggedright\arraybackslash}p{1.65cm}>{\raggedright\arraybackslash}p{3.25cm}@{}}
\toprule
Failure case & Why cleanup is missed & Sandman's handling \\
\midrule
Caller cancelled or timed out mid-create & the caller has left & provider finishes the create; pool keeps or deletes the late pod \\
API answers an error after storing the pod & error read as no pod & provider reads it by name; deletes it on a nonce match \\
Delete fails or its reply is lost & sent delete assumed done & provider retries until confirmed; owner reference after trainer pod deletion \\
\bottomrule
\end{tabular}
\end{table}

Command cleanup must also account for child processes. On timeout or reset, \execd{} kills the command's process
group and continues draining output while capping what it retains. It checks syntax before execution and accepts
an exit status only from a complete exit-status frame. The August build used in the campaigns lacked that status
check: a split frame could report a failed command as successful. The records do not establish how often this
happened (Appendix~\appref{app:g}).

Cancelled callers wait for their creates to finish, and late warm pods occupy capacity.
Commands are sent once because a timed-out command may have partially executed; retries would require
request deduplication in \execd{}. Sandman does not periodically sweep ownerless labelled pods, so this mechanism
does not cover unknown leak paths.

\subsection{Execution conditions for grading}
\label{sec:grading}

Our SWE editor reads the full file and checks its byte count and SHA-256 hash before writing, even when the model
sees only a shortened view. Files with encodings or line endings the editor cannot preserve require a shell edit
(Appendix~\appref{app:g}). The trainer extracts the patch against the task's base commit and excludes test-configuration
files. An earlier agent had used a \texttt{conftest.py} to bypass tests.

Each suite runs in a separate grading pod, fresh by default. The integration restores the official tests and declares
the image, network access, resource limits and other execution conditions. Before grading, an in-pod probe checks
outbound network access, CPU affinity, CPU quota, memory limit, user identity, time limit, hash seed and time zone
against the declared configuration. Undeclared or unreadable settings remain unverified. The image digest and test
suite are set separately. The trainer retries recognized infrastructure failures and resamples if the retries produce no verdict.
The grader first tries to apply the extracted patch. If application fails, it reverts the task image's pre-existing
edits and retries. Test failures and patches that still do not apply score zero. Section~\ref{sec:e2e} reports the functional validation.

A working sandbox can still host invalid tests. Before training, the integration checks that each task's reference
patch passes and an empty patch fails under the job's configuration. For two Verified tasks, the empty patch passed,
so the check rejected them. This preflight check costs 0.06--0.14 pod-hours per task. It grades stored patches
and does not test patch extraction. Appendix~\appref{app:h} gives the conditions, checks and historical results.

\subsection{Isolation and access control}
\label{sec:boundaries}

Earlier deployments blocked outbound traffic from some pods but left inbound endpoints unauthenticated. In a live check, an
unrelated pod invoked commands in outbound-guarded sandboxes. Such access to a grading pod could change its files
and reward.

\execd{} now requires a per-sandbox token on every call except health checks (Table~\ref{tab:bounds}).
Two \texttt{NET\_ADMIN} init containers install network rules that the \execd{} container cannot change. The outbound
guard restricts outgoing traffic and blocks access to the instance metadata service. The inbound guard restricts
access to \execd{}'s port to the trainer's pods and the node. In single-create tests, median creation time was
2.2\,s without the guards and 3.7\,s with both (Appendix~\appref{app:cmp}).

\begin{table}[t]
\caption{What each boundary protects and what it does not.}
\label{tab:bounds}
\centering\small
\begin{tabular}{@{}>{\raggedright\arraybackslash}p{1.5cm}>{\raggedright\arraybackslash}p{3.55cm}>{\raggedright\arraybackslash}p{2.35cm}@{}}
\toprule
Boundary & Protected & Not protected \\
\midrule
Execution & runc container, not privileged; guarded pods drop raw sockets & shared kernel; root user; no seccomp profile, user namespace or read-only root \\
Outbound & in-pod rules block all outgoing traffic but loopback and replies, failing closed & Pods without the outbound guard \\
Inbound & per-sandbox token on every call except health checks; in-pod allowlist of the node and consumers & past use (no request logs) \\
Metadata & the outbound guard & unguarded pods (metadata hop limit 2) \\
\bottomrule
\end{tabular}
\end{table}

In a separate live check, all 6 pods labelled egress-denied reached the internet because only a patched provider
installed the guard. The provider now refuses that label unless it installs the corresponding rules.

We use runc's shared kernel. A microVM per sandbox, using Firecracker~\cite{firecracker} or
Kata~\cite{kataarch}, requires bare metal or nested virtualization~\cite{ec2nested}.
gVisor~\cite{gvisor} only partly supports the packet filter used by our outbound guard~\cite{gvisorcompat};
we have not measured its cost. Pods set no seccomp profile, and the kubelet sets no default~\cite{kubeletconfig}. We do not claim protection against a malicious co-tenant.

\section{Evaluation}
\label{sec:eval}

\begin{figure*}[t]
\centering
\includegraphics[width=\textwidth]{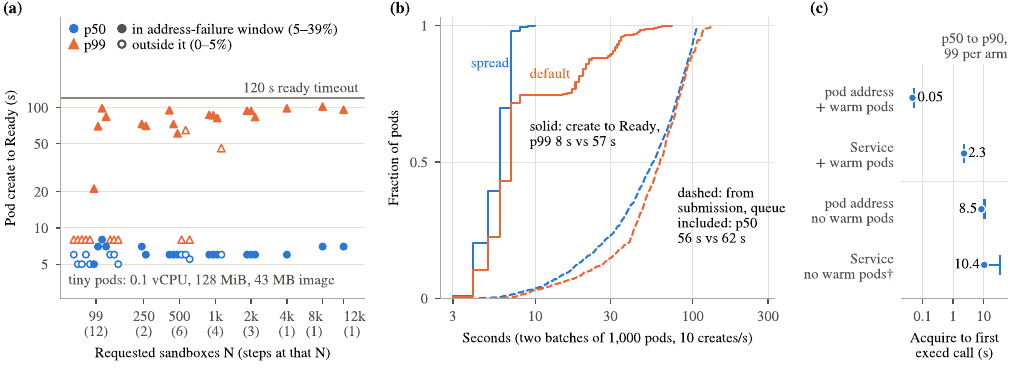}
\caption{Measurements of sandbox creation. (a)~Create-to-Ready p50 and p99 across all 30 ramp steps and
controls, using tiny pods and a cached image. Filled markers fall within the address-assignment failure window.
(b)~One sequential pair of 1,000-pod batches at 10 creates/s, showing create-to-Ready time and total wait from
submission, including the pacer queue. (c)~Acquisition to first successful \execd{} call, with 99 sandboxes per arm:
dot p50, whisker p90; $\dagger$ marks overlap with the failure window.}
\label{fig:scale}
\end{figure*}

The training runs use Qwen3.5-35B-A3B with GRPO in an asynchronous trainer. Most runs use 16 B200 GPUs: eight for
the learner and eight for SGLang inference. The 64-GPU campaign uses 32 GPUs for each. Unless stated otherwise, each
sandbox requests 2\,vCPUs and 8\,GiB of memory, with resource limits set to the same values.

Both arms of each A/B comparison ran concurrently on the shared cluster. Reported confidence
intervals are 95\% unless stated otherwise. The measurements cover different versions of Sandman;
Table~\ref{tab:ranwhere} identifies the mechanisms used in each study. Appendix~\appref{app:a} provides workload
configurations, code versions and statistical methods.

\begin{table}[t]
\caption{Measured versions and current production deployment (all 16 mechanisms: Table~\appref{tab:a-ranwhere}).
A--D: historical campaigns; RL studies: earlier controlled comparisons; Now: current defaults, with the outbound
guard set by the SWE templates. The short training validation is reported in \S\ref{sec:e2e}. p: in part; o: option.}
\label{tab:ranwhere}
\centering\small
\begin{tabular}{@{}>{\raggedright\arraybackslash}p{4.1cm}ccc@{}}
\toprule
Mechanism & A--D & RL studies & Now \\
\midrule
Direct calls; paced, spread, owned creates & -- & p & \checkmark \\
Outbound guard on every sandbox & C--D & \checkmark & \checkmark \\
Token, inbound guard, probe, task preflight & -- & p & \checkmark \\
Retries, readiness watch, split pools & -- & -- & \checkmark \\
Warm pool and checked reuse & -- & grading pods & o \\
\bottomrule
\end{tabular}
\end{table}

\subsection{Sandman in production}
\label{sec:evalprod}

From 3 to 17 October, Sandman created 3.13 million sandboxes for 122 training runs across four production campaigns,
A--D. These campaigns used earlier versions of Sandman (Table~\ref{tab:ranwhere}) and reached 27 simultaneous runs
on 432 B200 GPUs. A subsequent production scale-up reached 11,858 concurrent sandboxes (\S\ref{sec:evalcreate}).

From late July to mid-October, about ten users outside these campaigns ran at least 299 jobs using Sandman,
with up to 896 GPUs in a single job (Appendix~\appref{app:b}).

Using logged sandbox-path failures from campaigns A--C, we estimated availability at 91.16--99.73\% per agent-session start, counting retries separately.
The corresponding trajectory-level estimates were at least 96.69\% with one backend retry.
Unreachable admission webhooks caused 99.5\% of campaign C's acquisition failures.

In campaign D, 69.5\% of recorded starts obtained a pod. Twelve of 39 runs encountered admission incidents and
counted retries as new starts; the other 27 obtained pods for 99.7\% of starts.
Table~\appref{tab:b-avail} separates attempt-level and trajectory-level rates.

In a separate sample of 198 sandbox pods across users, CPU use averaged 10.5\% of requested CPU and memory use
averaged 1.9\% of requested memory. Both averages are weighted by observed pod-minutes.
For campaigns A--C, we estimated sandbox resource costs at 1.7--4.0\% of GPU costs using on-demand list prices.
The range depends on whether shared CPU-node costs are allocated by requested CPU or memory
(Appendices~\appref{app:a}--\appref{app:b}).

\subsection{Creation at scale}
\label{sec:evalcreate}

During the scale-up of production jobs, Sandman requested up to 12,000 sandboxes, paced at 20--25 creates/s.
Peak concurrency reached 11,858, and median create-to-Ready time stayed at 5--7\,s (Figure~\ref{fig:scale}a).
Each sandbox used 0.1\,vCPU, 128\,MiB of memory and a cached 43\,MB image. We stopped at 12,000 requests to preserve
subnet address headroom. Across the ramp and additional rate checks, none of the 38,377 creates encountered an
admission error. These measurements did not cover sustained creation and destruction.

During a 2.2-hour window, 19--25\% of pods in each ramp step failed their first IP assignment, and create-to-Ready
p99 reached 102\,s. All ramp steps above 99 sandboxes
fell within this window, so the ramp alone could not separate the effects of concurrency from changes in cluster
conditions.

Follow-up measurements found more address failures on nodes receiving many new pods. We compared two batches
of 1,000 sandboxes at 10 creates/s on 375 eligible nodes. We ran the spread batch first, then default scheduling
after three idle minutes.

Spreading placed at most 3 pods from the batch per node, versus 65 with default scheduling. No pod in the spread
batch failed its first IP assignment, compared with 254 under default scheduling. All pods eventually became Ready.
Create-to-Ready p99 was 8\,s with spreading and 57\,s without (Figure~\ref{fig:scale}b).

Including the pacing queue, median submission-to-Ready wait was 56.2\,s with spreading and 61.8\,s without.
Both batches had a 50\,s median pacing wait. This was one sequential pair, so cluster conditions may have changed
between batches (Appendix~\appref{app:c}).

\subsection{Comparison with open-source systems}
\label{sec:evalcmp}

\begin{figure*}[t]
\centering
\includegraphics[width=\textwidth]{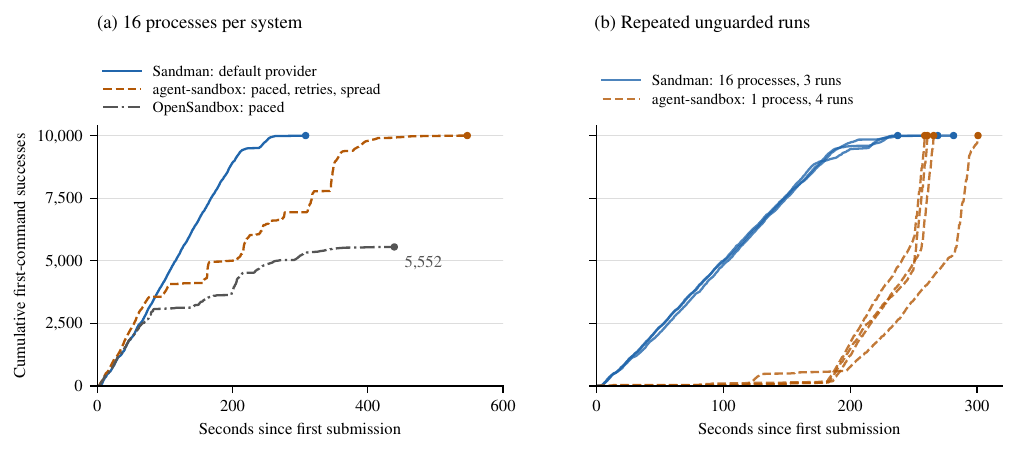}
\caption{Cumulative successful first commands from the start of each 10,000-request burst, including time in client
queues. Half and full delivery occur at the 5,000th and 10,000th success. (a)~One run per configuration, with
16 processes per system. Sandman uses its default inbound guard; agent-sandbox adds pacing, retries and soft
topology spreading; paced OpenSandbox reaches 5,552 successes. (b)~Repeated unguarded runs: 16 processes for Sandman
and one for agent-sandbox, with a curve for each run. Full-delivery times overlap. Settings: Appendix~\appref{app:cmp}.}
\label{fig:comparison}
\end{figure*}

We compared Sandman with three open-source systems: agent-sandbox~\cite{agentsandbox} and
OpenSandbox~\cite{opensandbox}, which create Kubernetes pods, and CubeSandbox~\cite{cubesandbox}, which restores
microVM snapshots. The pod-based systems ran one at a time on the same 40 dedicated 96-vCPU nodes.
CubeSandbox ran on 42 separate nodes of the same instance type with
nested virtualization. A common driver used each system's Python client. Sandman used its deployed provider;
Appendix~\appref{app:cmp} lists the versions and deployment settings.

\textit{Burst delivery.} Sixteen client processes submitted 10,000 requests together. The pod-based sandboxes used
cached images, 0.25\,vCPU and 1\,GiB of memory. We measured delivery from the first submission until each sandbox
successfully executed its first \texttt{true} command, including time in client queues. Each delivered sandbox
remained allocated for 60\,s before deletion.

Sandman's default provider, with its inbound guard enabled, and agent-sandbox with added pacing, retries and topology
spreading delivered all 10,000 sandboxes and left none at client exit. Half and full delivery took 116 and 308\,s
for Sandman, versus 198 and 547\,s for agent-sandbox (Figure~\ref{fig:comparison}a). Sandman allowed 20 creates/s per
process; agent-sandbox was capped at 53 creates/s total, matching Sandman's earlier observed rate.

Repeated runs without network guards used 16 processes for Sandman and one for agent-sandbox
(Figure~\ref{fig:comparison}b). Across three runs, Sandman delivered half in 100--104\,s and all in 237--282\,s.
Across four runs, agent-sandbox delivered half in 241--275\,s and all in 259--301\,s. Sandman delivered the first half
earlier, while full-delivery times overlapped. Appendix~\appref{app:cmp} also reports delivery with both guards enabled.

\begin{table}[t]
\caption{Cached-image bursts, one run per row. Each used 16 processes and 10,000 requests, except
OpenSandbox's partial 13-process run. Usable: first command succeeded; left: present at client exit.
Guards are off unless marked. CubeSandbox's guest probe was disabled.}
\label{tab:cmp}
\centering\small
\begin{tabular}{@{}>{\raggedright\arraybackslash}p{4.05cm}rr@{}}
\toprule
Client configuration & Usable / requested & Left \\
\midrule
Sandman, default (inbound) & 10,000 / 10,000 & 0 \\
agent-sandbox, default & 1,124 / 10,000 & 2,570 \\
\quad with our retries & 6,459 / 10,000 & 2,913 \\
\quad pacing and retries & 10,000 / 10,000 & 0 \\
\quad pacing, retries and spread & 10,000 / 10,000 & 0 \\
OpenSandbox, partial run & 2,016 / 8,125 & 1,292 \\
\quad with our pacing & 5,552 / 10,000 & 2,502 \\
CubeSandbox & 7,740 / 10,000 & 5 \\
\quad first command retried & 8,860 / 10,000 & 61 \\
\bottomrule
\end{tabular}
\end{table}

Agent-sandbox's unpaced client encountered API throttling and SDK errors (Table~\ref{tab:cmp}).
Retries alone did not recover all requests; configurations with pacing and retries
completed all 10,000. Paced OpenSandbox delivered 5,552 sandboxes, and its lifecycle server restarted twice. Earlier
Sandman configurations also failed requests at 100--200 creates/s; Appendix~\appref{app:cmp} reports these runs and
their cleanup outcomes.

CubeSandbox's highest-completion run delivered 8,943 of 9,843 requests, with a median delivery time of 33\,s
among successes. Its guest probe failed on our deployment. Disabling it left a readiness race.

We also submitted two 10,000-request bursts 15\,s apart, exceeding the nodes' approximately 11,600 pod slots.
With a 1,500\,s readiness budget, Sandman delivered all 20,000 requests within 951\,s of the first burst.

Serial creation was faster on agent-sandbox: median 1.45\,s versus Sandman's 2.18\,s and OpenSandbox's 2.46\,s
at 1\,vCPU and 2\,GiB. CubeSandbox restored its smaller 0.25-vCPU, 1-GiB template in 0.08\,s.

\textit{SWE images and tool calls.} Each pod-based system created 1,024 sandboxes with 2\,vCPUs and 8\,GiB of memory.
Each system used a different SWE image of compressed size 1.25--1.28\,GB, with uncached layers. Guards were disabled.
An init container copied agent-sandbox's daemon into the image.

Each run completed 30,720 subsequent no-op \texttt{true} calls. Median latency was 2.0\,ms for Sandman,
14.3\,ms for agent-sandbox and 203.5\,ms for OpenSandbox. The time to deliver half the sandboxes was 20.3, 33.3 and
25.4\,s, respectively; full delivery took 32.0--35.6\,s. All creates, commands and deletes succeeded
(Table~\appref{tab:cmp-swe}).

\textit{Cancellation and cleanup.} Each client started 300 creates, with cancellation scheduled at random times
within 10\,s. Sandman and agent-sandbox left no sandboxes at client exit. Sandman also left none with pacing
disabled, including cancellations after pods had become Ready. OpenSandbox left sandboxes running for 191 of 215
cancelled creates. This test used a four-hour lifetime rather than its ten-minute default.
CubeSandbox's synchronous client continued creating environments after cancellation.

Most configurations ran once. Sandman received more tuning, and our OpenSandbox and CubeSandbox deployments may be
suboptimal.

\subsection{Prewarming and reuse}
\label{sec:evalprep}

\emph{Image preparation.} In recorded re-gradings with 1.3--8.4\,GB SWE-Gym images, median creation time was
17.4\,s when the node needed to pull the image (422 pods), versus 4.1\,s when it was cached (230 pods). The same
pattern appeared when records were grouped separately by image, node, or half-hour window
(Figure~\appref{fig:prep}a; Table~\appref{tab:d-blocks}). In a separate preflight over 280 tasks, a 120\,s readiness
timeout that included image pulling left nine tasks without a grading verdict. \sys{} therefore makes this timeout
configurable per job.

\emph{Prewarming.} We compared three pairs of 16-GPU training runs, with 128 trajectories in flight per run.
Only grading-sandbox provisioning changed; rollout sandboxes were created on demand in both arms. The warm pool
served 93--96\% of gradings from ready pods. The per-run medians averaged 6.2\,s for on-demand grading acquisition
and 0.12\,s with the pool (Figure~\appref{fig:prep}b). With a 2,100\,s idle timeout, total reserved sandbox vCPU-hours
rose to 2.46 times the on-demand total, mostly because of idle warm pods.

Pool-on runs reached step 24 9.4\% sooner. In two pairs, however, most of that difference came from a first step
about 400\,s shorter, much larger than the grading wait removed by prewarming. Over steps 2--24, the mean
step-interval difference was $-8.7$\,s (pool on minus off; 95\% CI $-38.4$ to $+21.0$\,s). These three pairs do not
establish a steady-state training speedup from prewarming.

Trace replays with a 300\,s idle timeout estimated warm-hit rates of 94.5--97.7\%, with roughly half the idle
pod-hours of the 2,100\,s replay (Figure~\appref{fig:prep}c).

\emph{Checked reuse.} A separate three-pair training comparison found that checked reuse reduced pod creations per
grading by 29\% (95\% CI 22--36\%) relative to a single-use warm pool. Reused pods also waited idle between uses,
so fewer creates did not necessarily reduce pod-hours. In replays of eight recorded traces, creating grading pods
on demand used the fewest pod-hours (Appendix~\appref{app:e}).

Across the recorded studies, median state-check time was 0.23--0.51\,s. The checks required a fresh pod in 8 of
64,199 SWE-bench Verified checks, compared with 23.7\% of checks on SWE-Gym, where rebuilt artifacts often prevented reuse
(Appendix~\appref{app:i}).

\subsection{Sandboxes and training}
\label{sec:e2e}

\begin{figure}[t]
\centering
\includegraphics[width=\columnwidth]{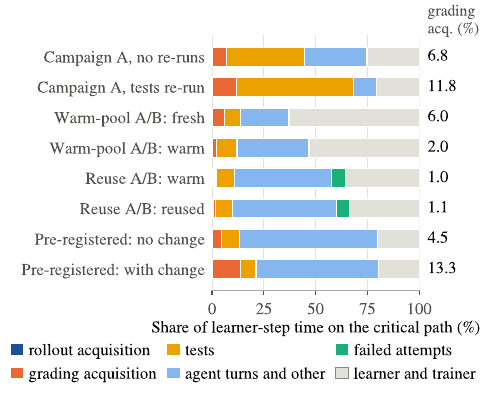}
\caption{Phases of the trajectory that ended each learner wait, across 3,734 steps. Campaign~A is production;
the rest are controlled. Campaign~A used up to two suites; two arms re-ran suites at batch consumption.
The reuse A/B's failed attempts fell within incident windows. Right-hand values give grading acquisition's share;
the pre-registered arms are with and without the grading change.}
\label{fig:critpath}
\end{figure}

\emph{Waiting for training data.} In campaigns A--C, waiting for training data accounted for 65--78\% of
learner-step time. To locate sandbox delays within these waits, we traced 3,734 learner steps from campaign A and
three controlled experiments. For each wait window, we identified the trajectory whose completion ended the wait
and measured the phases overlapping that window
(Figure~\ref{fig:critpath}; Appendix~\appref{app:critpath}).

Rollout-sandbox acquisition accounted for at most 0.9\% of learner-step time on this measured path. Grading acquisition
accounted for 4.5--6.8\% with fresh pods, 1--2\% with the warm pools in \S\ref{sec:evalprep}, and 13.3\% with the
capped pool described below. Test execution accounted for 8--11\% with one suite and 38\% with two.
Faster provisioning could also free rollout slots earlier and change which trajectories finish next, so the shares
do not predict end-to-end training speedup.

\emph{Combined grading changes.} Six pre-registered pairs of 150-step runs compared the original grading
integration with a combination of task preflight, pooled grading with checked reuse, and authentication on pooled
pods. The pool was capped at 32 pods and replenished at 0.3 creates/s. Under these limits, 50.9\% of gradings waited
for a new pod, with a median wait of 38.6\,s among those gradings.

At step 150, we evaluated eight samples per task on 112 held-out tasks. The modified arm's mean task-level pass
rate was 2.33 percentage points lower than the original integration's (95\% paired $t$-interval $-4.22$ to $-0.43$;
exact sign-flip $p=0.0625$). Preflight also excluded three training tasks. Because the task set, grading path and
authentication changed together, this experiment cannot attribute the result to any individual change
(Appendix~\appref{app:j}).

\emph{Validation of the deployed version.} A separate 20-step run checked sandbox creation, tool calls, patch
extraction, grading and cleanup with the deployed provider, daemon and grader. It used 23 SWE-Gym tasks,
16 B200 GPUs and 64 trajectories in flight. Sandboxes requested 2\,vCPUs and
4\,GiB; prewarming and reuse were disabled. The run completed 640 gradings without a patch-application failure and
left none of its 1,281 pods behind. In 165 gradings across six tasks, the grader reverted image-provided edits
before applying the extracted patch.

The run had no control arm or held-out learning endpoint (Appendix~\appref{app:current-validation}).

\section{Lessons from Operating Sandman}
\label{sec:lessons}

\emph{Budget for aggregate creation traffic.} Per-provider limits do not bound concurrent jobs' combined bursts.
An unpaced evaluation disrupted admission for our work and other jobs. We added pacing and increased the webhook
service from one replica to three. Reassess aggregate traffic and shared-service capacity as jobs are added.

\emph{Check placement and address capacity.} Our workers were restricted to 115 nodes already full of other jobs'
pods. The production scale-up consumed addresses in subnets shared with GPU nodes. Include node eligibility,
memory and pod addresses alongside CPU in capacity planning.

\emph{Disable unused Service links.} Namespace Services caused Kubernetes to inject about 23,400 environment
variables into new containers. Pod-address clients do not need them.

\emph{Retain cleanup responsibility.} Track accepted creates through cleanup after caller cancellation.
Owner references provide fallback when the trainer pod is deleted (\S\ref{sec:cancel}).

\emph{Measure training progress and cost.} Prewarming shortened grading acquisition and increased reserved CPU-hours.
Its steady-state training benefit remained inconclusive. We create on demand, with prewarming and reuse optional
(\S\ref{sec:evalprep}).

\emph{Validate extraction and tests.} Checking stored reference patches bypasses extraction from the agent's
workspace. Exercise extraction and patch application under the job's conditions, including image-provided edits
(\S\ref{sec:e2e}). Preflight task removal also changes the training distribution.

Appendix~\appref{app:incidents} documents 41 operational cases, causes, fixes and bounded costs. These records are
not a complete failure census.

\emph{Limitations.} We evaluated Sandman on one EKS cluster. Training studies use SWE-style Python tasks,
one harness and an asynchronous trainer. The production peak used small pods. Prewarming and reuse studies each
have three pairs. CubeSandbox used nested virtualization without its guest
probe. Replays fix trajectory timelines, incident-loss estimates are upper bounds, and security checks cover our own pods.
Paired training comparisons predate the current provider and daemon. The current version's short run validates
functionality without measuring a learning benefit.

\section{Related Work}
\label{sec:related}

\emph{Sandbox platforms.} DSec~\cite{dsec2026} manages multiple sandbox backends for agentic training. Kimi K2~\cite{kimik2}
uses Kubernetes, while AgentENV in Kimi K3~\cite{kimik3} uses Firecracker microVMs.
E2B, Modal and Daytona~\cite{e2b,modal,daytona} provide execution and lifecycle APIs.
The OpenHands Software Agent SDK~\cite{openhandssdk} integrates local and remote execution.
We compare agent-sandbox~\cite{agentsandbox}, OpenSandbox~\cite{opensandbox} and CubeSandbox~\cite{cubesandbox}
(\S\ref{sec:evalcmp}). Agent-sandbox also supports daemon injection into unmodified images.

\emph{Execution overhead and reuse.} The Rollout Infrastructure Tax~\cite{rollouttax} measures execution-substrate
overhead, excluding inference and reward computation, and recommends locality-aware warm pools. Our paired runs
measure grading acquisition, resource reservations and training effects. DeltaBox~\cite{deltabox} studies checkpoint and
rollback of filesystem and process state.
\sys{} restores a recorded baseline of grader-modified state (\S\ref{sec:sources}).

\emph{Training and grading.} AReaL, SkyRL-Agent and TideRL~\cite{areal,skyrlagent,tiderl} address asynchronous training
and multi-turn scheduling. GLM-5 and LEGO-RL~\cite{glm5,legorl} handle environment failures.
DAPO~\cite{dapo} filters identical-reward groups. Infrastructure-noise studies~\cite{anthropicnoise,epochswebench}
show that resource settings and image rebuilds can change coding-evaluation results.
SWE-Gym, SWE-smith and SWE-rebench~\cite{swegym,swesmith,swerebench} build and validate tasks.

\emph{Operational studies.} ML Productivity Goodput, sparing strategies, GUARD and
SAKURAONE~\cite{mlgoodput,sparing,guard,sakuraone} examine fleet efficiency, reliability and workload behavior.

\section{Conclusion}
\label{sec:conclusion}

\sys{} provides stateful environments for tool execution and evaluation in production agentic RL on shared Kubernetes.
The service handled millions of sandbox creations in our production campaigns. In the measured multi-client burst,
Sandman completed delivery sooner than tuned agent-sandbox. Tests on SWE images also found lower median no-op
command latency than both pod-based baselines.

Operating the service requires accounting for shared admission and address capacity, pacing requests and retaining
cleanup responsibility after cancellation. Faster provisioning must also be weighed against training progress and
reserved resources. Prewarming shortened grading acquisition but increased reserved CPU-hours, and the paired runs did not
establish a steady-state training speedup.
We will open-source the daemon, provider and pool.\label{sec:bodyend}

\bibliography{references}
\bibliographystyle{mlsys2025}
\section*{AI Usage Disclosure}
\iffinalsubmission\else\textcolor{red}{\textbf{[DRAFT --- the authors must confirm or correct every statement below before submission.]}}\fi

The authors led the research design, experiment planning and interpretation of results, and determined the paper's
organization and conclusions. They made the design and operational decisions and reviewed the code, analyses and
manuscript. Claude Code (Anthropic, using Claude models) and Codex (OpenAI) assisted with the activities described below.
\iffinalsubmission\else\TODO{add any other tools team members used}\fi

\textbf{Design and analysis.} The authors set the experimental objectives and approved the protocols and analysis
methods. AI tools helped critique candidate plans, including the cancellation-window tests using a stand-in Kubernetes
API and the analysis of recorded image-preparation measurements. The authors evaluated these suggestions and decided
which to adopt.

\textbf{Literature search.} AI tools searched for and summarized related work. Every cited work was checked to
exist, and the authors checked that each is represented correctly.

\textbf{Implementation and experiments.} AI assistance included drafting and revising code for the sandbox daemon,
provider, warm pool, trainer integration, and measurement, testing, analysis and figure scripts. The authors specified
the requirements and initial code structure, reviewed the code and evaluated it through tests and experiments.
AI agents also assisted with launching and monitoring jobs and collecting cluster measurements under the authors'
oversight. The authors approved operational changes.

\textbf{Data.} Production measurements and benchmark results came from recorded system executions. The evaluated
agent models generated trajectories as part of the experiments. Cancellation tests used simulated responses from a
stand-in Kubernetes API, and reuse tests used deliberately planted residue. These test inputs are described in the
corresponding experiments.

\textbf{Manuscript preparation.} The authors directed, reviewed and edited the paper, with AI assistance in drafting
and revising text across the main paper and appendix, preparing tables and figures, and assembling submission files.
They verified the claims against source records; every reported number is traceable to a source file in their
evidence ledger.

The authors take full responsibility for the correctness, originality and integrity of the submission.

\clearpage
\setcounter{section}{0}\setcounter{table}{0}\setcounter{figure}{0}
\renewcommand{\thesection}{\Alph{section}}\renewcommand{\thetable}{A\arabic{table}}\renewcommand{\thefigure}{A\arabic{figure}}
\twocolumn[{\centering\LARGE\bfseries Appendix\par}
\vskip 0.2in]
\newcolumntype{L}[1]{>{\raggedright\arraybackslash}p{#1}}
\newcommand{\hd}[1]{\begin{tabular}[b]{@{}c@{}}#1\end{tabular}}

This supplement gives the configurations, methods and qualifications needed to interpret the main paper's
measurements. It distinguishes production observations, controlled comparisons, trace replays and defect
reproductions. Section~\ref{app:incidents} lists the recorded operational cases;
Section~\ref{app:current-validation} reports a separate functional validation of the deployed version.
The symbol \S{} refers to the main paper.

\section{Workloads, implementations and endpoints}
\label{app:a}

Production campaigns comprise our training campaigns A--D and other users' jobs on \sys{}.
We analysed run logs, trajectory records, API audit logs and cluster metrics. Dedicated experiments shared
this cluster with other workloads; the same team operates the cluster. A/B arms ran concurrently unless stated
otherwise. Replays held recorded trajectory timelines fixed while running the real pool code in virtual time.
Reproductions test specific failure paths, not their production frequency. Attempts include retries; trajectories,
pods, gradings, labels and learner steps have separate denominators.

The RL runs used Qwen3.5-35B-A3B and GRPO, with groups of eight samples per task, usually four groups per
learner step. Tasks came from SWE-bench Verified~\cite{swebenchverified} and SWE-Gym~\cite{swegym}, with one image
per task. Unless stated otherwise, sandbox requests and limits were both 2\,vCPU and 8\,GiB (16\,GiB for a few
heavy repositories). Tests had a 1,800\,s limit; campaign creates had a 300\,s budget including image pulls.
The post-prompt token budget was 16,384, or 32,768 in two SWE-Gym runs. Inference ran outside the sandboxes.
Our trainer keeps a rollout sandbox until grading completes; the campaigns' trainer released it before grading.

\begin{table}[htbp]
\caption{Sandman's interfaces. The pool is optional; prewarming is disabled in our SWE default.}
\label{tab:api}
\centering\small
\begin{tabular}{@{}>{\raggedright\arraybackslash}p{1.9cm}>{\raggedright\arraybackslash}p{5.85cm}@{}}
\toprule
Component & Calls \\
\midrule
Pool (one Ray actor per job) & \texttt{acquire}(image, base commit, spec) $\rightarrow$ sandbox id, address; \texttt{release}(id); \texttt{prewarm}(images); \texttt{drain}() \\
Provider (library) & \texttt{create}(spec) $\rightarrow$ handle; \texttt{destroy}; \texttt{health} \\
\execd{} (HTTP, one per pod) & \texttt{execute\_command}, \texttt{read\_file}, \texttt{write\_file}, their byte forms, \texttt{env}, \texttt{reset}, \texttt{state}, \texttt{health} \\
\bottomrule
\end{tabular}
\end{table}

\emph{Code builds.} Table~\ref{tab:a-versions} maps the recorded source builds to workloads;
Table~\ref{tab:a-ranwhere} summarizes their mechanisms. The daemon build names denote specific changes:
initial reads at most 2,000 lines; syntax-check validates wrapped commands; whole-file reads complete files;
cleanup kills the original process group and drains output while capping retained bytes; strict-frame validates
exit-status frames; state-check adds baseline-check routes. The whole-file build with added state-check routes
still used the old exit-status parser. Harness builds were initial, checked-read, later checked-read and
guarded-editor. Checked reads verify length and SHA-256; the guarded editor refuses files it cannot preserve.
The available records do not provide a complete per-workload source-revision and image-digest manifest.

\begin{table*}[tp]
\caption{Code builds used by each workload. Build names identify the implementations in the records.
The provider and pool belong to \sys{}; the trainer and harness belong to the RL training stack.}
\label{tab:a-versions}
\centering\small
\begin{tabular}{@{}L{2.8cm}L{6.7cm}L{6.7cm}@{}}
\toprule
Workload & Provider, pool and trainer & Daemon and execution path \\
\midrule
Campaigns A--D & Runtime-image provider before cancellation shielding; no pool. C--D patched in the outbound guard and disabled Service variables; A--B did neither. & Whole-file build from a carrier image, with the old parser. The campaigns' trainer released rollout pods before grading. \\
Scale-up and placement & Initial provider-side cancellation fix plus run-specific resource names, optional affinity and spread; driver-side pacing. & Strict-frame build. \\
Warm-pool and reuse A/Bs & Campaign-C provider with the provider half of the cancellation fix and resource-matched warm handoff. Reuse added recycling, warm-budget accounting, pacing and state checks. & Whole-file build in both; reuse added state-check routes. Both retained the old parser and the campaigns' trainer. \\
Frozen deployment and pre-registered comparison & Frozen pool, reuse guard and task check; soft spread. Full cancellation fix on pooled grading pods, provider half elsewhere. & Token-supporting build in rollout and grading pods; token enabled only on pooled grading pods. Test re-execution kept the whole-file build. Agent calls used a local pod-address forwarder. \\
Earlier validation & Grading-path build with the mechanisms marked VR in Table~\ref{tab:a-ranwhere}; rollout provider from the pre-registered study. & Token-supporting rollout/grading daemon; whole-file build for test re-execution. The pre-registered trainer had one actor change and the same forwarder. \\
System comparison and current validation & Earlier comparisons used a driver subclass with a readiness watch, refusal retries and one shared API/daemon thread pool. Final comparisons and current validation used the deployed provider with split pools. & Earlier token-supporting build, later adding keep-alive, byte routes and TCP\_NODELAY; final runs used the deployed daemon. Section~\ref{app:current-validation} checks the deployed provider, daemon and grader in training. \\
Cancellation tests & Four implementations: before the fix, cancellation fix, error-path fix and an alternative implementation. & Stand-in API and real-cluster checks are distinguished in Section~\ref{app:f}. \\
Execution reproductions & Real agent tools on a development host. Harness builds: initial, checked-read, later checked-read and guarded-editor. & Daemon builds: initial, syntax-check, whole-file, cleanup, strict-frame and state-check. \\
\bottomrule
\end{tabular}
\end{table*}

The earlier validation checked grading mechanisms on 21 Verified tasks in one 20-step run. Its pool created
891 pods: 167 on demand and 724 warm. Of these, 441 were acquired and 450 never used. The 789 acquisition events
comprised 441 first uses and 348 recycled uses; 784 completed and five were unfinished at the stop.
Shadow-grading pods were counted separately. This validation is distinct from the current-version check.

The frozen deployment fixed its implementation and configuration before launch, including keeping newly found
bugs unfixed within that study. It checked pools of 230 and 280 SWE-Gym tasks and 115 Verified tasks, then ran
three 40-step SWE-Gym training jobs, each with 16 GPUs and 48 trajectories in flight. The Verified pool was used
only for task checks. Table~\ref{tab:a-frozen-revisions} gives the recorded revisions for this study; they do not
identify the later pre-registered comparison or current deployment.

\begin{table*}[tp]
\caption{Recorded revisions for the frozen deployment. Rollout pods retained a separate historical provider;
the trainer also used the recorded task-adapter and timeout overlays. These revisions alone are not a full
deployment manifest.}
\label{tab:a-frozen-revisions}
\centering\small
\begin{tabular}{@{}L{3.3cm}L{8.9cm}L{4.0cm}@{}}
\toprule
Component & Recorded revision & Scope \\
\midrule
Provider and pool & \nolinkurl{57e56468e678301e169ab298aabffc79b49e6f0b} & Grading pool and task checks \\
\execd{} & \nolinkurl{97baf17d74bc7cce94b2700744adc78903f98621} & Study-created pods \\
Trainer & \nolinkurl{213288ee5718cbe7c6519dc830847fddbcad4ce8}, plus overlays & SWE-Gym runs and task checks \\
\bottomrule
\end{tabular}
\end{table*}

\begin{table*}[tp]
\caption{Mechanisms used by each workload. A--B, C--D: production campaigns; WP, RU: the warm-pool and reuse A/Bs; PR: pre-registered
package arm; VR: earlier validation run; Now: current production defaults of our code and SWE job templates.
The short training validation has separate checks in Section~\ref{app:current-validation}. \checkmark{} ran; --
did not; o option, off by default; g grading pods only; p in part (Table~\ref{tab:a-versions}).}
\label{tab:a-ranwhere}
\centering\small\setlength{\tabcolsep}{2.1pt}
\begin{tabular}{@{}lccccccc@{}}
\toprule
Mechanism & A--B & C--D & WP & RU & PR & VR & Now \\
\midrule
Agent calls by pod address, not a Service & -- & -- & -- & -- & p & p & \checkmark \\
Service environment variables off & -- & \checkmark & \checkmark & \checkmark & \checkmark & \checkmark & \checkmark \\
Soft topology spread over nodes & -- & -- & -- & -- & \checkmark & \checkmark & \checkmark \\
Paced creates & -- & -- & -- & g & g & g & \checkmark \\
Outbound guard on every sandbox & -- & \checkmark & \checkmark & \checkmark & \checkmark & \checkmark & \checkmark \\
\execd{} token and inbound guard & -- & -- & -- & -- & p & g & \checkmark \\
Create retries, readiness watch, split thread pools & -- & -- & -- & -- & -- & -- & \checkmark \\
\execd{} keep-alive and byte routes & -- & -- & -- & -- & -- & -- & \checkmark \\
Warm pool & -- & -- & g & g & g & g & o \\
State-checked reuse & -- & -- & -- & g & g & g & o \\
In-pod configuration probe & -- & -- & -- & -- & -- & g & \checkmark \\
Reference/empty-patch check of the task set & -- & -- & -- & \checkmark & \checkmark & \checkmark & \checkmark \\
Create keeps an owner on cancel & --$^\dagger$ & --$^\dagger$ & p & p & p & p & \checkmark \\
Strict exit-status parsing & -- & -- & -- & -- & p & p & \checkmark \\
Test-configuration files left out of patches & -- & -- & -- & -- & -- & -- & \checkmark \\
Recognized infrastructure failures resampled & \checkmark & \checkmark & \checkmark & \checkmark & \checkmark & \checkmark & \checkmark \\
\bottomrule
\end{tabular}
\\[2pt]{\scriptsize $^\dagger$Campaigns cancelled creates only at test re-run deadlines (not measured).}
\end{table*}

\emph{Endpoints and uncertainty.} Delivery tests count a sandbox as usable when its first command succeeds.
Create time runs from the provider's API request to a healthy response; create-to-Ready uses Kubernetes timestamps
at one-second resolution. Acquisition includes the pacing queue. A--C availability is estimated from logged
sandbox-path failures, as defined in Section~\ref{app:b}; it is not a count of confirmed successful acquisitions.
Pod lifetime is creation to deletion, including idle time. Reserved vCPU-hours
multiply pod lifetime by the CPU request; they do not measure CPU execution. Unless specified otherwise, intervals
are two-sided 95\% $t$-intervals over run pairs (three for the A/Bs, six for the pre-registered comparison).
Single-proportion intervals use Wilson's method; trajectories are not treated as independent A/B replicates.

\section{Production scope and availability}
\label{app:b}

\begin{table*}[tp]
\caption{Production cohorts and availability. A--C report estimated failure complements per attempt and then per
request with one backend retry (Table~\ref{tab:b-availability-counts}); neither is derived from the pod-count columns.
D counts starts obtaining a pod, rather than a working sandbox. Spans are campaign windows; counting cutoffs differ as described below.
A--C ran 50 learner steps, D 300. A trajectory can use two grading pods.}
\label{tab:a-campaigns}\label{tab:b-avail}
\centering\small
\begin{tabular}{@{}lrrrrrrr@{}}
\toprule
 & Tasks & Runs $\times$ GPUs & Span (h) & Rollout pods & Grading pods & Attempts & Availability \\
\midrule
A & 24 & $32\times16$ & 30.1 & 79,861 & 113,407 & 81,102 & 99.73\% / 99.98\% \\
B & 24 & $16\times16$ & 10.6 & 40,136 & 59,164 & 42,409 & 96.98\% / 99.65\% \\
C & 88 & $35\times64$ & 57.3 & 396,810 & 568,930 & 453,155 & 91.16\% / $\geq$96.69\% \\
D & 88 & $39\times16$ & 82.7 & 720,332 & 1,147,595 & 1,036,028 & 69.5\% got a pod \\
\bottomrule
\end{tabular}
\end{table*}

A--C use $1-F/N$ for per-attempt availability: $N$ counts first attempts, coordinator retries and backend retries;
$F$ counts logged sandbox-path failures. The trajectory-level estimate is $1-L/R$, where $R$ counts first requests
and coordinator retries, and $L$ counts second backend attempts that also failed (Table~\ref{tab:b-availability-counts}).
The attempt and request denominators come from result JSONs cut at the last learner step; failure tallies cover
whole logs. For C, whole-log starts number 454,562 versus 453,155 in the step-cutoff JSONs, a difference of about 0.31\%.
These failure complements therefore do not give matched successful-acquisition counts or unique pod counts.
C had 448 coordinator retries, which could recover some second-backend failures; its $\geq$96.69\% marks this
conservative request-level estimate, not a statistical confidence bound.

\begin{table}[htbp]
\caption{Counts behind the A--C availability estimates. $N$: attempts; $F$: sandbox-path failures;
$R$: requests; $L$: second-backend failures. Denominators use step-cutoff JSONs; failures use whole logs.}
\label{tab:b-availability-counts}
\centering\small
\begin{tabular}{@{}lrrrr@{}}
\toprule
Campaign & $N$ & $F$ & $R$ & $L$ \\
\midrule
A & 81,102 & 218 & 80,903 & 18 \\
B & 42,409 & 1,282 & 41,275 & 144 \\
C & 453,155 & 40,039 & 427,239 & 14,122 \\
\bottomrule
\end{tabular}
\end{table}

Campaign C's 40,039 sandbox-path failure warnings included 39,834 HTTP-500 pod-create failures
($39{,}834/40{,}039\approx99.488\%$, rounded to 99.5\%), 203 Ready timeouts, one evicted pod and one unreachable \execd{}.
The warnings retained only the first error line. Sampled full tracebacks identified failed calls
to unreachable Kyverno and Kueue mutating admission webhooks. The HTTP-500 count is exact;
its webhook attribution comes from those sampled tracebacks, not an individual traceback match for every failure.

Campaign A had 27 held-out tasks; D removed sphinx-8475 at launch and reached 27 concurrent runs (432 GPUs).
The D extension comprised 48 planned training configurations, attempted twice with individual relaunches;
its incident counts refer to distinct job executions (Section~\ref{app:incidents}). A--C's reported rollout
counts are nonmissing pod-creation timings in trajectory records that started by the last learner step and had
ended (\texttt{by\_end} true and \texttt{t\_end} present). This metric does not count unique pod UIDs or all successful
session starts. Removing the step cutoff adds 0.010--0.013\% to the same creation metric; it does not explain the
difference from the failure complements. Without a start-to-pod reconciliation, the remaining difference is
unclassified. The 122 campaign runs used about
21,407 B200-hours. Other users ran at least 299 sandbox jobs, from about ten owners, with up to 896 GPUs per job.
In D, 12 of 39 runs encountered admission incidents; retries inflated the attempt count. The other 27 acquired
495,566 of 497,170 requested pods (99.7\%). Thus the aggregate includes our operational failures.

Campaign A averaged 1.8 creates/s but reached 132 in one second; 79.8\% of within-run image re-requests arrived
within ten seconds. A 198-pod, 14-day resource sample, weighted by pod-minutes, used 10.5\% of requested CPU and
1.9\% of requested memory. These are utilization samples, not measurements of every campaign pod.
At list prices of \$160.645/h for eight B200 GPUs and \$4.752/h for 96\,vCPU and 192\,GiB, summed campaign trajectory
time gives 3.30--3.75 sandbox-hours per GPU-hour. CPU and memory allocation estimates are 1.6--1.8\% and 3.3--3.7\%
of GPU cost; allowing 2--7\% extra for test re-runs and evaluation gives the paper's 1.7--4.0\% range.
These are allocation estimates under the campaign trainer's sandbox lifetime, not billed marginal cluster cost.
The inputs come from the AWS EC2 Price List for ap-south-1, offer version \texttt{20261008184850}, effective
2026-10-01: On-Demand Linux, shared tenancy, no preinstalled software, capacity status Used.
The recorded rates were \$160.64525/h for p6-b200.48xlarge (SKU \texttt{QUSTXWNCN8VJ5E2X}) and
\$4.752/h for c5d.24xlarge (SKU \texttt{MSSXBMQZ8JAEGDFC}); the GPU rate is shown rounded above.

\section{Production scale-up, placement and routing}
\label{app:c}

The production scale-up covered 30 measured cells, including rate checks, from 99 to 12,000 requested
sandboxes on 260 CPU nodes. It used the provider-side cancellation fix, a strict-frame daemon and driver pacing, with a cached 43\,MB image,
0.1\,vCPU and 128\,MiB per pod, and no network guards. Creates were paced at 20--25/s outside the
5--200/s rate sweep; deletes at twice the create rate. The production deployment reached 11,858 concurrent sandboxes.
At 12,000 requests, 11,826 acquisitions completed within budget, 91 failed at the provider's Ready deadline and
83 exhausted the 600\,s acquisition budget. Over the cell, 48 timed-out creates later became healthy and were
retained warm. The late-arrival total and concurrent peak have different observation times. Every step ended
with zero leftovers. Across the scale-up and additional rate checks, none of the 38,377
creates encountered an admission error. Shared subnet headroom set the stopping point: about 33.6k addresses were
free before the scale-up and 26.5k after teardown, with zero scale-up pods remaining. Warm prefixes stayed attached to nodes;
the latter reading is not a count at peak concurrency.

All scale-up steps above 99 fell within a 2.2-hour period with 19--25\% initial pod-address failures. Ready latency
had 5--7\,s medians and a maximum step p99 of 102\,s. Concurrency and cluster state therefore cannot be separated.
A later follow-up stopped its second 5,000-pod step under the co-tenant guard. Later spread runs had fewer
address failures, but ran 1.5--3.5 hours after the unspread runs.

The controlled placement check ran one sequential pair of 1,000-pod cells on 375 nodes, both at 10 creates/s:
spread first, 180\,s idle, then default placement. Spread used all 375 nodes (at most three pods/node), versus
72 nodes (at most 65); zero versus 254 pods first failed address assignment. All became Ready; Ready p99 was
8 versus 57\,s. Spread pulled the image on 263 nodes while default placement was fully cached. We stopped after
two of twelve pre-registered cells under the co-tenant failure rule. This is one non-randomized pair, not a pacing
comparison. In the separate 99-pod routing cells, warm Service access first succeeded at 2.28\,s versus 0.05\,s by
pod address; every warm Service access needed a retry, while DNS itself took 4--9\,ms.

\begin{table*}[tp]
\caption{Capacity observations and configuration scopes. Settings govern separate budgets; a pool cap is not
a job-wide sandbox limit. The placement result is one sequential pair on the shared cluster.}
\label{tab:c-limits}\label{tab:configuration}
\centering\small
\begin{tabular}{@{}L{2.5cm}L{6.5cm}L{7.2cm}@{}}
\toprule
Constraint & Setting and scope & Observation or limit \\
\midrule
Create/API pressure & Current provider: 20 creates/s, 12 API threads and up to 30 refusal re-sends through the pacer. API and daemon calls use separate pools. & One provider budget covers pool creates, warm builds and replacements. Independent providers and direct fallbacks add traffic; no budget spans jobs. Earlier 100--200/s configurations failed requests (Section~\ref{app:cmp}). \\
Node eligibility & Select the CPU-pool label rather than one instance type. & Pinning restricted workers to 115 full nodes despite capacity elsewhere. All nine affected jobs completed after recreation with pool-label selection. \\
Placement & Soft spread over hostnames, maximum skew 2, \texttt{ScheduleAnyway}; the scheduler may override it. & At 10 creates/s, the two 1,000-pod batches placed at most 3 versus 65 pods/node. Initial IP failures were 0 versus 254; Ready p99 was 8 versus 57\,s. All became Ready. Spread ran first, so changing cluster conditions remain a confound. \\
Pod addresses & VPC CNI: /28 prefix delegation, warm-IP target 2 and minimum-IP target 16 per node. These targets are not exact free-address counts. & During the scale-up, 19--25\% of pods first failed address assignment within a 2.2-hour window. Free addresses were about 33.6k before the scale-up and 26.5k after teardown, with zero scale-up pods; warm prefixes remained attached. The scale-up stopped at 12,000 requests to preserve headroom. \\
Pool capacity & Optional cap counts idle, creating and checked-out pool pods. Reuse A/B: 96 pods; pre-registered arm: 32, including 16 warm, replenished at 0.3/s. Reuse limit: eight. & Direct fallback bypasses the cap and pool pacer. Prewarming is off by default; these study-specific caps do not bound all job pods. \\
Readiness & Campaign create budget: 300\,s including image pulls. Scale checks used a 120\,s Ready deadline and a 600\,s acquisition budget; the two-burst comparison used 1,500\,s. & Acquisition also includes the pacing queue. A late warm arrival is not an on-time acquisition. Budgets are configurable per job, not a single system-wide timeout. \\
Routing & Direct pod addresses; Service environment-variable injection disabled. & An unpaced 1,000-Service run had 241 failures; the paced run created all 1,000. In separate 99-pod warm cells, first-command access took a median 2.28\,s through Services and 0.05\,s by pod address. \\
\bottomrule
\end{tabular}
\end{table*}

\section{Comparison with other sandbox managers}
\label{app:cmp}

\emph{Environment.} We tested agent-sandbox 1.0.8, OpenSandbox 1.2.0 and Cube 0.7.3. The pod-based systems ran
sequentially on the same 40 m8id.24xlarge nodes (96\,vCPU, 384\,GiB, maximum 300 pods/node), with about 11,600
available pod slots. Cube used 42 separate nodes of the same type, including its control plane, with nested
virtualization; only two bare-metal nodes were available, so we did not calibrate its overhead. Cube used
0.25\,vCPU, uncapped, and 1\,GiB. Its template's guest readiness probe was disabled after it reported FAILED;
a readiness race remained. It retained its built-in private-address filter.

All cached-burst sandboxes used 0.25\,vCPU and 1\,GiB. The burst used python:3.12-slim for \sys{} and OpenSandbox,
and agent-sandbox's python-runtime-sandbox:1.0.8
with its non-root, default-seccomp configuration and one-second readiness probe. Unless marked, network guards
were disabled: \sys{} still required its token, agent-sandbox used unmanaged network policy, and OpenSandbox
had no egress sidecar. A common Python SDK driver started the client processes at one clock, ran \texttt{true},
held each sandbox for 60\,s, then deleted it. Half-delivery is the 5,000th first-command success; full-delivery time
is reported only when all requested sandboxes succeeded. Leftovers were counted at client exit, before our cleanup.
These are delivery measurements, not training-throughput measurements.

\begin{table*}[tp]
\caption{Cached 10,000-request bursts. Values after a slash are attempted requests when a planned driver did not
start. Times are seconds; -- means full delivery was not achieved. Repeated rows give observed ranges, not intervals.}
\label{tab:cmp-bursts}
\centering\small
\begin{tabular}{@{}L{8.2cm}rrrr@{}}
\toprule
System and configuration & Usable & Half & All & Left \\
\midrule
\sys{}, current default (inbound guard): 16 processes, 20/s each & 10,000 & 116 & 308 & 0 \\
\sys{}, 16 processes, 10/s each; three repeats & 10,000 & 100--104 & 237--282 & 0 \\
\sys{}, both guards, 16 processes, 10/s each & 10,000 & 154 & 315 & 0 \\
agent-sandbox, 16 processes, default & 1,124 & -- & -- & 2,570 \\
agent-sandbox, 16 processes, retries & 6,459 & -- & -- & 2,913 \\
agent-sandbox, 16 processes, retries and 3.3/s each & 10,000 & 224 & 576 & 0 \\
agent-sandbox, same pacing/retries plus soft spread & 10,000 & 198 & 547 & 0 \\
agent-sandbox, one process; four repeats & 10,000 & 241--275 & 259--301 & 0 \\
OpenSandbox, one process & 562 & -- & -- & 9,435 \\
OpenSandbox, 13 of 16 processes started & 2,016/8,125 & -- & -- & 1,292 \\
OpenSandbox, 16 processes, paced at 3.3/s each & 5,552 & -- & -- & 2,502 \\
Cube, 16 processes & 7,740 & -- & -- & 5 \\
Cube, 16 processes, retry first command for 10\,s & 8,860 & -- & -- & 61 \\
Cube, 63 of 64 processes started & 8,943/9,843 & -- & -- & 19 \\
\bottomrule
\end{tabular}
\end{table*}

\emph{Tuning and failures.} The final \sys{} default used 12 API threads per process, readiness watches, separate
API/daemon thread pools and soft spread. Its 30 allowed refusal retries were unused. The tuned agent-sandbox
budget of about 53 creates/s matched an earlier achieved \sys{} rate, not the final default's 320/s configured
budget. Its 44\% full-delivery difference therefore compares complete configurations, not one mechanism.
Most configurations ran once, and we spent more effort tuning \sys{}; these are not proven optimal settings.
The repeated full-delivery ranges overlap. Two one-process agent-sandbox repeats used driver-named claims and two
SDK-generated names. OpenSandbox's paced run included two lifecycle-manager restarts.

Earlier \sys{} configurations also failed: one process at 100/s delivered 9,998/10,000; 16 at 12.5/s without
retries delivered 4,087/10,000; a 15-of-16-process launch delivered 3,225/9,375; and 16 at 6.25/s with eight attempts
delivered 8,339/10,000. Refusals were server, webhook or throttling errors before storage; all created \sys{} pods
were deleted. These four configurations enabled both guards; the first two also used the training namespace's three
admission mutations. Earlier drivers used one shared API/daemon thread pool. Splitting it removed queued health work,
but concurrent retry, pacing and code changes prevent assigning the final delivery improvement to that change alone.
Earlier neighboring probes shared \sys{}'s service account and API fairness flow; later probes used their own.
Those before/after measurements do not isolate effects on other tenants.

Two 10,000-request bursts 15\,s apart delivered all 20,000 with \sys{} in 951\,s under a 1,500\,s readiness budget.
The one-process, driver-named-claim agent-sandbox run stalled at 2,351 after 65 minutes and left 17,649 claims.
We did not isolate the cause or repeat it with paced or SDK-generated claims; it does not implicate default rollback.

\emph{Serial creation.} One client repeatedly created a sandbox, ran one command and deleted it for 300\,s.
Creation latency was recorded separately from the following first command. At 1\,vCPU/2\,GiB, median creation
latency was 1.45\,s for agent-sandbox, 2.18\,s for \sys{} without guards (3.71\,s with both) and 2.46\,s for
OpenSandbox. Cube took 0.08\,s with its 0.25-vCPU, 1-GiB template. First-command times were 16, 2.8, 206
and 151\,ms for agent-sandbox, \sys{}, OpenSandbox and Cube, respectively. Cube needed a retry for 123 of 603
first commands; retries every 0.1\,s succeeded within 0.33\,s of creation. Thus these creation medians and the
burst tests' first-command delivery times use different endpoints. \sys{}'s readiness probe first ran 1\,s after
container startup, then every 2\,s.

\begin{table}[htbp]
\caption{One 1,024-sandbox SWE-image burst per system, with newly introduced image layers. All sandboxes, 30,720
subsequent commands and deletes succeeded. Latencies are seconds for delivery and milliseconds for subsequent calls.}
\label{tab:cmp-swe}
\centering\small
\begin{tabular}{@{}lrrrr@{}}
\toprule
 & Half & All & Call p50 & Call p99 \\
\midrule
\sys{} & 20.3 & 33.0 & 2.0 & 9.0 \\
OpenSandbox & 25.4 & 32.0 & 203.5 & 222.0 \\
agent-sandbox & 33.3 & 35.6 & 14.3 & 64.2 \\
\bottomrule
\end{tabular}
\end{table}

The SWE bursts used 16 processes, 2\,vCPU/8\,GiB pods, first \texttt{true} plus 30 subsequent no-ops and a 60\,s hold.
The systems used three different Verified task images, 1,283, 1,268 and 1,252\,MB compressed, whose layers
overlapped neither earlier benchmark pulls nor one another.
Guards were disabled. \sys{} paced creates at 10/s per process; agent-sandbox used retries without pacing.
agent-sandbox retained its restricted security configuration and copied its own staging v1.0.8 portable daemon
through an init container; preparing the task and init-container images encountered pull-QPS backoffs.
These calls did not execute an agent trajectory or a test suite.

Cancellation tests requested 300 creates per client and cancelled uniformly within ten seconds. \sys{} left zero
pods (254 cancellations in the default run; 221 unpaced); agent-sandbox's default SDK naming and rollback left zero
(210 cancellations). OpenSandbox left 191 after 215 cancellations, with a configured four-hour lifetime rather
than its ten-minute default. Cube's synchronous creates continued after cancellation: all 19 cancelled creates
remained, plus one failed delete. These tests assess cancellation under these client configurations, not expiry.

\section{Image preparation and warm capacity}
\label{app:d}\label{app:e}

The image-preparation analysis used SWE-Gym re-grading records for 59 images from a 230-task set, on 184 nodes.
These records measure provider create time, excluding the first command. ``Pulled'' means the image reference
was absent, as inferred from Kubernetes events; layers could still be cached. The 1.3--8.4\,GB sizes are whole
compressed images, not transferred bytes. The 422 pulled records had median provider create time 17.4\,s, versus
4.11\,s for 230 present records. Blocking by image, node or half-hour preserved a positive difference
(Table~\ref{tab:d-blocks}). Events have one-second resolution. Another 359 pods lacked grading records and had
longer pulls, so the observed cold tail is optimistic; download and unpacking cannot be separated.

\begin{table}[htbp]
\caption{Differences between pulled and present median provider create times within blocks containing both cache
states. Cache state was observed, not randomized. Differences are seconds.}
\label{tab:d-blocks}
\centering\small
\begin{tabular}{@{}lrrr@{}}
\toprule
Block & Count & Median difference & Range \\
\midrule
Same image & 39 & 18.2 & 6.0--140.6 \\
Same node & 101 & 14.1 & 4.7--81.3 \\
Same half-hour & 6 & 12.4 & 10.1--16.7 \\
\bottomrule
\end{tabular}
\end{table}

\begin{figure*}[!htbp]
\centering
\includegraphics[width=\textwidth]{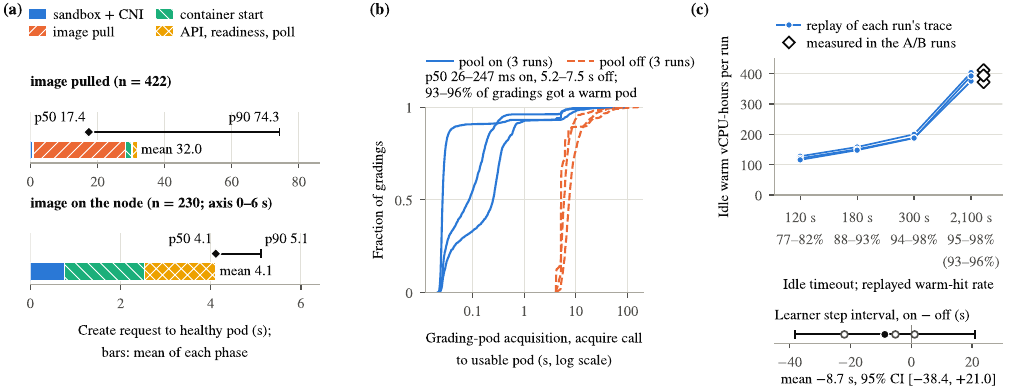}
\caption{Creation phases, acquisition savings and prewarming costs.
(a)~Phase means sum to mean create-to-healthy time for SWE-Gym images. Cache state was observed in controlled-run
records, not assigned; axes differ, and markers show p50 and p90.
(b)~Grading-pod acquisition with the pool off and on, across three controlled pairs.
(c)~Replayed idle warm vCPU-hours and hit rates by idle timeout; diamonds mark measurements, with hit rates in
parentheses. Below: measured step-interval difference with 95\% CI.}
\label{fig:prep}
\end{figure*}

\emph{Warm-pool A/B.} Three concurrent pairs used campaign C's recipe on 16 GPUs with 128 trajectories in flight;
only grading provisioning changed. They ran 25 steps on 88 tasks (14,513 trajectories). The pool kept two pods
per key with a 2,100\,s idle timeout. The recorded frozen pool implementation keys interchangeable pods by requested
image reference and base commit; ``image key'' refers to the image component, not interchangeability across commits.
Resource requirements were checked at handoff. Pair one ran four days before pairs two and three with identical manifests;
the later API was slower. Averaging per-run medians, acquisition fell from 6.177 to 0.119\,s, with 93--96\% hits.
Table~\ref{tab:e-warm} reports each run's total reserved sandbox vCPU-hours, including rollout and grading pods.
Let $H_{\mathrm{on},i}$ and $H_{\mathrm{off},i}$ denote those totals for pair $i$. The main paper's 2.46 is the
equal-weight mean of the three paired ratios, $\frac{1}{3}\sum_i H_{\mathrm{on},i}/H_{\mathrm{off},i}\approx2.4613$,
computed from unrounded measurements. Pooling the printed hours before division gives
$2{,}024/825\approx2.453$, which rounds to 2.45. Thus 2.46 is the average paired total-cost ratio, not the ratio of
all pool-on hours to all pool-off hours. Both estimators include idle reservations and do not measure CPU execution.
Two pairs had first steps about 400\,s shorter, and wall time
to step 24 fell 9.4\% (ratio 0.906, CI 0.870--0.941). Steady step intervals changed by $-8.7$\,s (CI $-38.4$ to
$+21.0$), which is inconclusive. Learner work was itself about 20\,s shorter; that is not a direct sandbox effect.

\begin{table}[htbp]
\caption{Total reserved sandbox vCPU-hours per training run in the warm-pool A/B, including rollout and grading
pods. Each row is one concurrent pair. Hours are rounded as recorded in the study summary.}
\label{tab:e-warm}
\centering\small
\begin{tabular}{@{}lrr@{}}
\toprule
Pair & Pool off & Pool on \\
\midrule
1 & 281 & 681 \\
2 & 249 & 652 \\
3 & 295 & 691 \\
\midrule
Sum & 825 & 2,024 \\
\bottomrule
\end{tabular}
\end{table}

\emph{Checked-reuse A/B.} Three pairs ran 40 steps on 86 checked tasks (33,656 trajectories), comparing a
single-use warm pool with at most eight reuses. Both used a 96-pod budget, at least two pods/key and a 300\,s idle
timeout. The paired creates-per-grading ratio was 0.710 (CI 0.640--0.780); step intervals changed by $+3.0$\,s
(CI $-15.5$ to $+21.6$). Shadows agreed on 3,642/3,646 reused gradings and 2,103/2,103 fresh pairs. Three disagreements
came from a pod with an unknown pytest fault and one from a flaky test. No task was known to be reuse-sensitive,
so this A/B does not establish that checking was necessary for these tasks. Both A/Bs used the old status parser:
it cannot change measured acquisition latency or create counts, but may affect agent behavior.

\emph{Replay limits.} Replaying each trace with three seeds, a 300\,s idle timeout retained 94.5--97.7\% hits at
48--50\% of the 2,100\,s replay's idle hours. Replayed baseline hits exceeded measured hits by 1.8--3.5 points.
Across eight workload traces, on-demand single-use had the fewest pod-hours. Reuse could reduce creates but raised
pod-hours 2--60\% relative to single-use warming. These replays fix the trajectory timeline, omit pacing and use
one pod specification; they are capacity-policy comparisons, not estimates of learning speedup.

\section{Cancellation and execution failure checks}
\label{app:f}\label{app:g}

The cancellation harness ran the unmodified provider and pool against a stand-in Kubernetes API with controlled
responses and gates. The final suite had 25 windows, 20 repetitions, four versions and Python 3.11/3.12: 4,000
trials, including 1,000 on the final fix, which left no pod without an owner.
Every trial included a sentinel pod; none of the 1,000 final-version trials deleted it.
Table~\ref{tab:f-windows} groups the boundaries by failure mode.
The stand-in does not model admission, scheduling, address allocation, image pulls, finalizers, contention or
production frequency. Twenty-five real-cluster trials (16 cancellation, nine error-path) left zero owned pods;
six error-path trials injected a client response. A separate owner-reference check removed an orphan about
11\,s after its owner was deleted. Queued work can still create after cancellation, and late grading pods can remain
warm. Retry state is lost if its event loop exits; owner garbage collection is the fallback.

\begin{table*}[tp]
\caption{Grouped cancellation windows and execution reproductions. Cancellation rows cover all 25 windows;
execution rows test concrete defects and fixes, not occurrence rates. The heredoc row includes syntax-check latency.}
\label{tab:f-windows}\label{tab:g-heredoc}
\centering\small
\begin{tabular}{@{}L{3.0cm}L{6.2cm}L{7.0cm}@{}}
\toprule
Check & Before or intermediate behavior & Final check and remaining limit \\
\midrule
Cancel/timeout/drain (windows 1--15) & Stored creates could lose their owner; the cancellation fix retained ownership. & Owned late results were deleted or retained warm by policy; cancellation does not prevent every queued create. \\
Stored create error (16--18) & The cancellation fix handled the reset case but leaked on two 504 paths. & Nonce read-back and owned cleanup covered all three paths. \\
Failed/lost delete (19--24) & The cancellation fix cleaned only the applied-delete case; other paths could leak. & Owned deletes retry; actor exit can still lose pending retries. \\
Name conflict (25) & An alternative delete-after-any-error handler deleted the other owner's pod in 40/40 trials. & Ownership checks preserved the conflicting pod in all 40 window-25 trials. \\
Child processes & Old timeout/reset left up to two children. & Seven of eight scenarios cleaned fully. A child in its own session survives group reset; state checking detects it and restore kills it. \\
File editing & A 2,000-line read cap made 408 edits discard 655,315 lines; 408 other edits were refused. & Complete-read stacks made 816/816 exact edits. Exact-match tests do not cover approximate matching, which can rewrite other text. \\
Bytes and line endings & Intermediate checked-read code corrupted invalid UTF-8 or mixed endings. & Current guards reject unsupported inputs; supported LF, CRLF and BOM content is preserved. \\
Exit status & The whole-file build reported success for 200/200 pinned failing commands; an intermediate fix still accepted empty status. & Strict framing returns the status or restarts the shell, losing shell state. Historical campaign frequency is unknown. \\
Unterminated heredoc & Old code waited 300\,s and returned 124. & Syntax checking rejected observed cases in about 4\,ms; valid-command overhead was about 2\,ms (1.12 to 3.30\,ms). \\
Output flood & An endless stream hit a 2\,GiB limit and OOM-killed old code after 9.63\,s. & Capped retention with continued draining reached the normal 20.01\,s timeout after draining 5,264\,MiB. \\
\bottomrule
\end{tabular}
\end{table*}

Execution reproductions used fresh django-11099 containers on one 8-vCPU, 30-GiB development host, the real agent
tools over loopback, and daemon source mounted read-only as PID 1; the image was unmodified. They establish
pre/post behavior under these inputs, not robustness under every command or shared-cluster load. Containers still
share a kernel. Current network guards and tokens do not make this a hostile-tenant security evaluation.

\section{Grading conditions and task checks}
\label{app:h}

We graded all 500 Verified reference patches twice in each cell of a $2^{8-4}$ resolution-IV design, four times
in the reference configuration, and the empty patch twice in four cells. The eight factors were CPU quota
(uncapped/one CPU), affinity (all/one core), memory limit (four times/equal to request), egress (open/denied),
user (root/uid 1000), shared memory (64\,MiB/1\,GiB), hash seed (unset/zero), and time zone (image UTC/one of ten
pinned rules). Cells requested 1\,vCPU and 2\,GiB, or 4\,GiB for matplotlib/xarray; the separate dataset-reference
configuration used 2\,vCPU/8\,GiB, root, open egress and no explicit seed or zone. With base factors Q, K, M, N,
generators were U=KMN, S=QMN, P=QKM, T=QKN. Main effects alias with interactions of order three or higher;
two-factor effects remain grouped into seven alias chains. Fisher tests used Holm correction over eight main
effects or seven chains at $\alpha=0.05$; we manually checked every flip against logs.

There were 481 stable tasks, 15 condition-dependent tasks (11 egress, four CPU quota/affinity), two consistent
image-provenance failures and two flaky tasks. No campaign trained a condition-dependent Verified task.
The separate 2,401-task SWE-Gym census found 115 condition-dependent tasks (4.79\%, CI 4.01--5.72), excluding seven
with discordant reference repeats and twenty with other differences. Four visible CPU cases may be repeat noise;
excluding them gives 4.62\% (CI 3.85--5.54). Twenty-two modin tasks OOM-failed at 8\,GiB but passed at 16\,GiB.
Reference-patch checks do not cover every agent-patch pattern: we also observed user-dependent grading outcomes
for agent patches that installed stub packages.

The official suites for django-7530 and django-10097 accepted empty patches. The reference/empty check rejected
both. On campaign D's consumed positives for these two tasks, 19.0--22.9\% were wrong under restored or hardened
tests~\cite{sweabs}, depending on the suite, representing 0.31--0.37\% of all D labels. A test-configuration patch
scan covered A--B and four runs each of C--D: 211,625 trajectories and 72,496 distinct patches. Ten consumed
positives matched; inspection found no verdict-changing edit. The other 31 C and 35 D runs were not
restored from cold storage, so this is incomplete coverage, not a production-wide absence result.

Task checks require the reference patch to pass twice and the empty patch to fail twice on fresh pods,
and on reused pods when enabled,
stopping early on failure. They cost 0.06--0.14 pod-hours/task. In a 280-task pool, nine tasks produced no verdict
under a 120\,s readiness budget that included pulling; the budget is now per job. This procedure does not validate
undeclared conditions or prevent extraction through every agent-edited file.

The in-pod configuration probe was exercised on 240 fresh SWE-Gym pods. Deliberately wrong timeout and hash-seed
declarations were refused on all 240; this does not validate every wrong field. Across 7,514 stored readings, 116 from 85 pods
found a 16-versus-8\,GiB declaration mismatch. All 789 acquisitions in the earlier validation passed the probe.
Unreadable fields remain unverified. The current-version validation is a separate cohort.

\section{State-checked reuse}
\label{app:i}

The reuse check covers repository state, hooks/configuration, installed-package metadata and path files,
selected system/home/temporary files, environment, time zone and processes. Its baseline is observed state, not
an image-integrity proof. Without deep mode it excludes installed files below depth two and shared memory;
it does not check directory modes, extended attributes or ACLs. A failed restore or recheck retires the pod.
Baseline capture took 0.4--0.8\,s and at most 64\,MiB; median checks took 0.23--0.51\,s off the acquisition path.
Fresh fallback occurred in 8/64,199 Verified checks but 593/2,505 SWE-Gym checks, mostly for artifacts or
package-metadata changes.

A three-arm test used 787 gradings per arm on 86 instances, combining real patches with planted residue.
Plain reset produced 64 wrong outcomes, including all 60 post-poisoning cases; checked reuse produced zero.
The one-sided upper bound is 3.4\% per instance, not per grading. These tests do not establish complete state
coverage. Nor can restoration fix a bad baseline: one pod with an unknown pytest fault was deleted before inspection.
The outcome rule retires a pod after no verdict or a result that contradicts an earlier verdict for the same
patch and test suite on any pod. It retired two of three faulty pods early but also three healthy pods
after flaky outcomes, and missed faults without a previous result. In a 69,812-grading replay, it prevented three
wrong reused labels at 0.12\% extra SWE-Gym creates; a stricter rule retired 135 healthy pods and added 18.7\%.

\section{Critical-path attribution}
\label{app:critpath}

The trace analysis covered 3,734 steps: 1,568 from A, 144 warm-pool, 234 reuse and 1,788 pre-registered steps.
Waiting for data occupied 78.0\% of step time in A, 73.6\% in B and 64.7\% in C; B and C use the result files'
learner-idle share. These give the main paper's rounded 65--78\% range.
For each learner wait, the analysis follows the selected graded groups backward through trajectory phases and failed
predecessors, including variance-based selection. Phase sums match waits within 0.003\,s, with 0--0.2\% unresolved.
The pre-registered runs lacked selection timing and used a two-window rule. Calibration kept total sandbox share
within $-0.8$ to $+0.3$ points, but understated the test component by 18--34\%. B--D have only per-run totals.
These are observed critical-path contributions, not predicted speedups: removing one delay can expose another
or free capacity for a different trajectory. In the warm-pool A/B, 41\% of pool-off steps did not wait and generation
was blocked by a full channel 25\% of the time. At three pairs, the minimum detectable interval change was about
39\,s (24\% of the 160.7\,s baseline); the inconclusive steady-step result is not evidence of equivalence.

Table~\ref{tab:cp-split} separates ordinary grading from test re-execution during batch preparation before
hand-over to the learner. Ordinary grading includes an initial second suite when used. The main paper's
4.5--6.8\% fresh-pod acquisition range excludes re-execution. Its rounded test-execution shares are 8--11\% for
one suite and 38\% for campaign A without re-execution, which used up to two suites. In the two re-execution arms,
batch hand-over re-executions contributed 9.5 of their 11.8 acquisition
points and 41.5 of their 56.5 test points. These are phase components within the pooled arms, not differences
from the no-re-execution row.
Rollout acquisition contributed at most 0.9\%, in the single-use warm arm of the reuse A/B.

\begin{table}[htbp]
\caption{Grading acquisition and test execution as percentages of total learner-step time.
The outside-incident row is a subset of the two reuse arms.}
\label{tab:cp-split}
\centering\small
\begin{tabular}{@{}L{4.6cm}rr@{}}
\toprule
Workload & Acquisition & Tests \\
\midrule
Campaign A, no re-execution & 6.8 & 37.9 \\
Campaign A, re-execution arms (pooled) & 11.8 & 56.5 \\
Warm-pool A/B, fresh & 6.0 & 7.8 \\
Warm-pool A/B, warm & 2.0 & 10.0 \\
Reuse A/B, single-use warm & 1.0 & 9.1 \\
Reuse A/B, checked reuse & 1.1 & 8.2 \\
Reuse A/B, outside incidents & 1.3 & 11.2 \\
Pre-registered, fresh & 4.5 & 8.4 \\
Pre-registered, capped reuse & 13.3 & 7.6 \\
\bottomrule
\end{tabular}
\end{table}

\section{Pre-registered learning comparison}
\label{app:j}

Before the first training launch, including the pilot, we fixed the arms, endpoint, seeds, analysis, interpretations
and stop rules in a committed protocol and an immutable cluster object. Deviations were append-only. The trainer's
verdict code was byte-identical to campaigns C--D. The package combined three changes: an 86-task checked set
rather than 89 tasks, checked warm reuse rather than fresh direct grading, and tokens on pooled grading pods.
The check rejected the two vacuous Django tasks and sphinx-8475 under denied egress, and admitted all 112 held-out
tasks. Prompt inputs were linked to its record by a SHA-256 chain. Both arms denied outbound traffic on every pod.
The pool cap was 32 (16 warm), replenishment 0.3/s and reuse limit eight; 2,361 gradings (4.3\%) used untokened
direct fallbacks. Every reused grading and 30\% of never-used gradings received a fresh shadow.

Each primary run used 16 GPUs, 64 in-flight trajectories and four groups of eight per step for 150 steps; a pilot
pair ran 20. All launched pairs were included, with a minimum of four. Pairs five and six could launch only if
both runs started by the pre-registered deadline, determined from the clock and launcher; both launched before
any evaluation was read. Each step-150 model was evaluated once on 112 held-out tasks, eight samples each at
temperature 1.0, all graded on fresh pods. The endpoint averages passes per valid sample over tasks, then takes
paired run differences. A 95\% paired $t$-interval wholly above/below zero meant ``package higher''/``no package
higher''; otherwise ``no difference detected''. Equivalence required the interval to lie within $\pm$3 points.

\begin{table}[htbp]
\caption{Held-out resolve rates. The package was lower in five pairs and tied in one. Rates are rounded to four
decimals; percentage-point differences are computed from the unrounded rates before rounding to two decimals.}
\label{tab:j-pairs}
\centering\small
\begin{tabular}{@{}lrrr@{}}
\toprule
Pair & Package & No package & Difference \\
\midrule
1 & 0.4286 & 0.4810 & $-5.25$ \\
2 & 0.4721 & 0.4721 & 0.00 \\
3 & 0.5112 & 0.5223 & $-1.12$ \\
4 & 0.4487 & 0.4810 & $-3.24$ \\
5 & 0.4531 & 0.4766 & $-2.34$ \\
6 & 0.4531 & 0.4732 & $-2.01$ \\
\midrule
Mean & 0.4611 & 0.4844 & $-2.33$ \\
\bottomrule
\end{tabular}
\end{table}

The mean difference was $-2.33$ points (SD 1.81; CI $-4.22$ to $-0.43$), giving the registered ``no package higher''
reading, not equivalence. The exact sign-flip $p=0.0625$ cannot be smaller with five non-zero pairs. Minimum-set
and five-pair intervals included zero; adding pairs was not outcome-driven. Because task set, grading path and
authentication changed together, the design cannot explain the lower score. Of the control's 25,246 consumed graded
labels, 623 came from the excluded tasks (2.5\%). The denominator counts consumed graded labels, rather than
the 28,800 learner records. The arms therefore trained on different data.

The package used 1.21 creates/grading and had higher acquisition latency: median/mean 7.8/20.3\,s versus 5.5/7.2\,s.
Under its cap and pacer, 50.9\% waited for a new pod (median 38.6\,s). Its 938 grading pod-hours and the control's
323 use different accounting methods and are not comparable. Four shadow disagreements involved flaky tests,
including one on a fresh pod. Deviations were concurrent pilot/primary launches without a free-capacity gate,
two shadow-integration fixes that changed no training run, one first-step communication-hang relaunch, and two
evaluation retries after GPU reset reports. The co-tenancy guard counted all our sandbox pods, including shadows,
and peaked at 901 of the approved 1,000; the training runs' own pods peaked at 712.

\FloatBarrier
\section{Full incident list}
\label{app:incidents}

These 41 cases come from notes, logs, audits and probes, not a representative failure census. Unless evidence
is specified, response verification was not recorded; unresolved causes and unfixed cases are explicit.
Only ten cases have recorded GPU-hour upper bounds (Table~\ref{tab:app-inc-gpuh}); the other cases have no
GPU-hour record. ``The extension'' consists of 48 planned training configurations for campaign~D, attempted
twice with individual relaunches. Its incident counts refer to distinct job executions, not unique configurations.

\subsection{Admission and addresses}
\label{tab:app-inc-capacity}

\medskip
\noindent \textbf{Managed service throttled session churn} (before \sys{}). Synchronized session recycling hit rate limits and shared
storage-partition throttling. Stop calls reached a reported spike of about 117/s against a 30/s limit, including
retries that stopped the same session 8--13 times; start and invoke limits had already been raised.
This motivated our own pods and pacing, but does not validate the replacement.

\medskip
\noindent \textbf{Admission webhook collapse}. Unpaced creates disrupted the webhook and other jobs. The cluster change record documents a move
from one 384-MiB replica to three 1-GiB replicas on separate nodes with a disruption budget; we also paced creates.
The later production scale-up had no admission errors in 38,377 creates. The sandbox role could not query the replica count
during that check, and load, pacing and pod specifications differed, so the observation does not isolate the change's effect.

\medskip
\noindent \textbf{Webhook outages in campaign~C}. Unreachable webhooks caused create failures; fresh-session retries and high availability lacked
remeasurement under the same load.

\medskip
\noindent \textbf{Scheduling tail in campaign~B}. Pending pods stalled eight runs, with no established cause or fix.

\medskip
\noindent \textbf{Pod-address assignment tail}. Spreading reduced address failures in one sequential controlled pair, but changing cluster conditions
left causality unresolved.

\subsection{Shared capacity}
\label{tab:app-inc-cotenancy}

\medskip
\noindent \textbf{Extension run failed during the production scale-up}. One run in campaign D's extension failed during the scale-up; other workloads' address failures
also rose during heavy-create periods, but the run's failure mechanism was not established. We capped later steps,
and a controlled guard aborted another co-tenant failure surge.

\medskip
\noindent \textbf{Node-pool pinning}. Instance-type pinning stranded nine jobs despite spare capacity; CPU-pool labels let all nine recreated
production jobs complete.

\medskip
\noindent \textbf{Event flood stalled the campaign driver}. A longer event-listing timeout restored driver decisions under namespace-event overload.

\subsection{Services and images}
\label{tab:app-inc-services}

\medskip
\noindent \textbf{Service environment injection}. A namespace snapshot contained 3,255 Services, including 3,187 per-agent Services from another fleet,
corresponding to about 23,400 Service-link environment variables per new container. This is an estimate from
the Service inventory, not a count from every campaign container. Default Service links amplified our per-agent
Service design; disabling them reduced command latency in a production smoke test.

\medskip
\noindent \textbf{Synchronized-start stall}. A Service burst coincided with API/DNS faults; staggered starts and pod-IP routing preceded a restart
without the pattern, but the cause remained unresolved.

\medskip
\noindent \textbf{Service routing failed under churn}. Direct IPs worked while Services failed; a controlled third attempt with IP forwarding had no
transport failures.

\medskip
\noindent \textbf{Unpaced Service creates throttled}. Pacing Service creates with pods resolved API priority/fairness throttling in the controlled retry;
no GPUs were involved.

\medskip
\noindent \textbf{Possible Service-name collision} (another team). Suspected name sharing prompted unique actor identifiers; cause and fix remained
unverified, impact unknown.

\medskip
\noindent \textbf{Readiness budget included image pulls}. Per-job readiness budgets replaced a budget exceeded by cold pulls; a later hold-out fully confirmed
only 7 of 9 affected tasks.

\medskip
\noindent \textbf{Base image expired from the registry}. An expired, unpinned base image prevented creation in earlier tool-use benchmark runs,
before the measured SWE campaigns. After rebuilding, digest pinning and longer registry retention, the next run
created sandboxes. Pinning alone does not prevent registry expiry.

\medskip
\noindent \textbf{Container store fell back to the root disk}. Boot-time NVMe assembly failed; we required assembly before runtime/kubelet startup.

\subsection{Execution and access}
\label{tab:app-inc-execution}

\medskip
\noindent \textbf{Commands that never returned}. A fail-open \texttt{bash -n} check refused malformed wrapped inputs in reproduction, replacing silent
stalls with shell errors.

\medskip
\noindent \textbf{Leftover child processes wedged sandboxes}. Timeouts left children while health checks accepted wedged pods; process-group cleanup and execution
probing preceded fewer hangs in separate production runs.

\medskip
\noindent \textbf{kubectl silently truncated results}. Default SPDY truncated reads yet reported success; WebSocket completed reproduction reads and reports
dropped streams, with no counted downstream effect.

\medskip
\noindent \textbf{Per-task working directory overridden} (another team's evaluation). Job-level workspace values overrode task values; the provider reversed precedence.

\medskip
\noindent \textbf{Tool-call transport failures in a cluster episode}. Tool/create failures spared direct-IP grading in the checked-reuse A/B; cause was unknown and no fix applied.

\medskip
\noindent \textbf{Rollout actors exceeded agent slots}. Matching trainer concurrency to actor slots unblocked rollout dispatch in the next run.

\medskip
\noindent \textbf{Unauthenticated inbound access to \execd{}}. Our own-pod probe crossed nodes without credentials. Tokens/ingress then passed controlled probes;
all 891 earlier-validation pool-created grading pods had the token and both guards. Campaigns had no inbound protection;
exploitation was neither observed nor audited.

\medskip
\noindent \textbf{Unguarded pods could get instance-metadata tokens}. Hop limit 2 exposed unguarded pods; controlled guards blocked metadata-token requests. A--B were exposed,
C--D were not; other unguarded workloads remain exposed and the hop limit is unchanged. Credential access was
deliberately untested.

\subsection{Grading}
\label{tab:app-inc-grading}

\medskip
\noindent \textbf{Planted test file passed a non-fix}. Excluding agent-planted test configuration prevented the reproduced non-fix pass in our trainer;
the campaigns' grader did not exclude it.

\medskip
\noindent \textbf{Grader cut test runs at 300\,s}. Omitted timeouts used the 300\,s daemon default; controlled 1,800\,s re-grading restored reference passes,
and probes now refuse the omission. Verified campaigns already sent 1,800\,s.

\medskip
\noindent \textbf{Egress-denied label without the guard}. All six pods in a live reference/empty-patch check reached the internet despite their deny labels:
only campaigns C--D's patched provider installed the guard. These were probe pods; all 2,540 deny-labelled pods
already in the namespace had the guard. The provider now installs and requires the guard before labelling
egress denied; controlled TCP/DNS checks passed.

\medskip
\noindent \textbf{Agent-backend error bursts in campaign~D}. Sampled backend errors coincided with webhook failures without established cause; the trainer retries
such sessions and never assigns a zero score.

\medskip
\noindent \textbf{Job controller lagged under one Job per grading}. Our one-Job-per-grading design delayed shared-controller repetitions; harvesting, time-to-live and caps
were checked only qualitatively.

\subsection{Reuse}
\label{tab:app-inc-reuse}

\medskip
\noindent \textbf{Sandbox reset deleted the task repository}. Preserving task repositories for git reset restored reuse in reproduction after workspace wipes had
forced fresh pods.

\medskip
\noindent \textbf{Warm pods built without the requested resources}. Missing prewarm resource requests were caught before measured runs; matching requirements and refusing
spec-free prewarms passed controlled checks.

\medskip
\noindent \textbf{Warm budget discarded recycled pods}. Replacement builds discarded returning recycled pods; revised in-check accounting and trimming passed
only unit tests, without cluster remeasurement.

\medskip
\noindent \textbf{State check missed edits to files that the image ships modified}. Baseline fingerprints and byte restoration fixed status-only checks and wrong restoration in two-image
reproductions; patch-apply failures had not changed labels.

\medskip
\noindent \textbf{A defective never-used pod was reused three times}. Four wrong zeros followed reuse; cause remained unknown after pod removal. Retirement for missing or
contradictory same-patch/test verdicts caught 2/3 defective pods but falsely retired three healthy ones in replay
(69,812 gradings).

\subsection{Cleanup}
\label{tab:app-inc-cleanup}

\medskip
\noindent \textbf{Pods left behind at job teardown}. We added owner references, central drain and a failed-actor reaper; the 122-run counters
never invoked the reaper and excluded create/delete error paths.

\medskip
\noindent \textbf{Cancelled creates leaked pods}. Cancellation bypassed error cleanup and prompted duplicate creates. In the stress test,
job logs and API audits recorded 13,516 rollout pods cumulatively: the manager deleted 7,612, and 5,904 remained
for garbage collection at teardown. Allowing up to 1,723 legitimately actor-held pods among the latter leaves
at least $5{,}904-1{,}723=4{,}181$ excess pods near job end. The 5,920 live-pod peak was a separate, earlier observation;
5,904 is not a count of pods individually classified as leaked. Owned cleanup preceded a leak-free production scale-up under changed conditions,
without establishing all-window coverage. A timeout-only workaround had failed.

\medskip
\noindent \textbf{Fallback creates bypassed the pod cap}. Direct-create fallback bypasses the cap; the guard deleted the job, and this limitation remains unfixed (\S4.2).

\medskip
\noindent \textbf{Pool not drained at job end}. An undrained detached actor left pods until owner-reference collection about 10 minutes later;
no fix was applied.

\subsection{Cluster events}
\label{tab:app-inc-cluster}

\medskip
\noindent \textbf{Node-group rolling updates under running jobs}. During rolling updates, 19 then 8 extension jobs were lost (9 of the first 19 had node-loss signatures);
we applied no fix.

\medskip
\noindent \textbf{Fleet-wide stall in the extension}. Ten runs were deleted at the 100-minute stall timeout amid shared I/O and cluster activity; cause remained unknown,
restrictions unverified, and the stall unattributed to the sandbox path.

\medskip
\noindent \textbf{GPU-side and trainer-side events}. GPU/trainer deadlocks, reset requests, collective hangs and file-system/memory failures prompted timeouts
and node avoidance; no verification or GPU-hour total was recorded.

\paragraph{Cost bounds.}
Table~\ref{tab:app-inc-gpuh} preserves the ten recorded bounds and their cause classes. Lifetime bounds charge whole
lifetimes, queueing included, only for jobs that died or never started. Allocated bounds charge failure-window GPUs,
with the lower end weighted by failed share; stall bounds charge GPUs over completed runs' excess time.
Each job contributes to one case.
The extension bounds cover 67 affected training-job executions: 29 in the synchronized-start case, 19 and 8 in
the two rolling updates, 10 in the fleet-wide stall and one during the production scale-up. They also include two
8-GPU evaluation jobs, one in the synchronized-start case and one in the first rolling update. The classifier assigns
each distinct execution to one case using mutually exclusive rules. These counts are not a census of all launched jobs.
Cause assessments distinguish \sys{} code, design and integration choices, unmodified Kubernetes defaults,
cluster or shared-service events, and cases whose cause was not established. The two path groups describe the
observed failure boundary; they do not establish causation or prove the absence of a sandbox contribution.

Allocated bounds use about 784 B200s for 1.46\,h at 84.9--85.4\% failure (webhook collapse), or 528 for 2.55\,h at
72.3\% (cancelled creates). Scheduling stall uses $8\times16\times(4.46-3.08)$ B200-h; campaign-C webhook stall is
24 noise-adjusted or 39 raw B200-h. Pinning covers six training runs and three evaluations, with zero lower bound.
Synchronized stall covers 29 runs that died or never started and one 8-GPU evaluation, all ended in the window; rolling
updates contribute $1{,}155+154$ B200-h. Service-link stall was 123--135 B200-h.

\begin{table*}[tp]
\caption{The 10 cases with GPU-hour upper bounds (B200-h), with cause assessments defined above. Ranges span the bounding methods used for each
case. The lower end is weighted by the failed share, adjusted for noise, or 0 for jobs that never started.
Each job's loss is assigned to one case, so rows can be summed. No other case has a GPU-hour record.}
\label{tab:app-inc-gpuh}
\centering\small\setlength{\tabcolsep}{4pt}
\begin{tabular}{@{}L{5cm}L{2.5cm}L{4.7cm}lr@{}}
\toprule
Case & Workload & Cause assessment & Basis & B200-h \\
\midrule
\multicolumn{5}{@{}l}{\emph{Sandbox path (Sections~\ref{tab:app-inc-capacity}--\ref{tab:app-inc-cleanup})}} \\
Cancelled creates leaked pods & stress test & sandbox implementation & allocated & 974--1,347 \\
Node-pool pinning & campaign~D & sandbox implementation & lifetime & 0--122 \\
Service environment injection & campaign~A & implementation and Kubernetes default & stall & 123--135 \\
Admission webhook collapse & evaluation & cluster or shared service & allocated & 973--1,145 \\
Webhook outages & campaign~C & cluster or shared service & stall & 24--39 \\
Extension run failed during the production scale-up & extension of D & not established & lifetime & 23 \\
Synchronized-start stall & extension of D & not established & lifetime & 1,170 \\
Scheduling tail & campaign~B & not established & stall & 177 \\
\midrule
\multicolumn{5}{@{}l}{\emph{Not attributed to the sandbox path (Section~\ref{tab:app-inc-cluster})}} \\
Node-group rolling updates & extension of D & cluster or shared service & lifetime & 1,309 \\
Fleet-wide stall & extension of D & not established & lifetime & 2,351 \\
\midrule
\multicolumn{4}{@{}L{14.8cm}}{Sandbox path: implementation 974--1,469; with a Kubernetes default 123--135; cluster or shared service 997--1,184; not established 1,370} & 3,464--4,158 \\
\multicolumn{4}{@{}L{14.8cm}}{Sandbox path by workload: stress test 974--1,347; evaluation 973--1,145; extension of D 1,193; campaigns A--D 324--473} & \\
\multicolumn{4}{@{}L{14.8cm}}{Not attributed to the sandbox path: cluster or shared service 1,309; not established 2,351} & 3,660 \\
\multicolumn{4}{@{}L{14.8cm}}{All 10 cases} & 7,124--7,818 \\
\multicolumn{4}{@{}L{14.8cm}}{In the campaigns (of their 21,407 B200-h): A's Service variables, B's scheduling tail, C's webhook outages, D's pin} & 324--473 \\
\bottomrule
\end{tabular}
\end{table*}

\section{Short training validation with the deployed defaults}
\label{app:current-validation}

A short training run exercised the deployed provider, daemon and grader through a complete training
workload. It used Qwen3.5-35B-A3B with GRPO, 16 B200 GPUs, 64 trajectories in
flight and 23 SWE-Gym tasks. Sandboxes requested 2 vCPUs and 4 GiB, with equal limits. Prewarming and state-checked
reuse were off. The run completed 20 learner steps and 640 gradings (Table~\ref{tab:current-validation-results}).

Patch extraction can include tracked edits already present in a task image. If the extracted patch does not apply
in the grading pod, the grader computes the image's own edits, reverts them and tries the patch again. A patch that
applies directly is graded without this recovery step. The run exercised recovery in 165 gradings across six tasks
and completed all 640 gradings without a patch-application failure. It created 1,281 sandbox pods and left none
behind.

The verifier reported 16 passed checks, one missing record and no failed checks. The missing item was the
structured task-admission record; the training log showed that all 23 tasks were admitted.

\begin{table}[!htbp]
\caption{Functional validation of the deployed version over 20 learner steps and 640 gradings.
Resolved trajectories are reported as successes / gradings.}
\label{tab:current-validation-results}
\centering\small
\begin{tabular}{@{}L{6.6cm}r@{}}
\toprule
Measurement & Result \\
\midrule
Patch-application failures (of 640 gradings) & 0 \\
Gradings that first reverted image-provided edits & 165 \\
Resolved trajectories on the six tasks with image-provided edits & 72 / 168 \\
Resolved trajectories on the other 17 tasks & 311 / 472 \\
Pods created & 1,281 \\
Pods left behind & 0 \\
Verifier checks passed & 16 \\
Verifier records missing & 1 \\
Verifier checks failed & 0 \\
\bottomrule
\end{tabular}
\end{table}

The run exercised creation, tool calls, patch extraction, grading and cleanup. The reference/empty-patch preflight
grades stored patches and does not exercise extraction. The run had no control arm or held-out learning endpoint,
so its resolved counts do not establish a learning benefit. It did not independently test every
cancellation path or security against a malicious tenant.

\end{document}